\documentclass[preprint]{elsarticle}

\usepackage{hyperref}
\usepackage{epstopdf}
\DeclareGraphicsExtensions{.eps}

\usepackage{graphicx}
\usepackage{amsmath}
\usepackage{mathptmx}
\usepackage{pstricks}
\usepackage{graphicx}
\usepackage{longtable}

\newcommand{\bfr}{{\mathbf{r}}}

\newcommand{\bfE}{{\mathbf{E}}}

\newcommand{\bfe}{{\mathbf{e}}}

\newcommand{\bfk}{{\mathbf{k}}}

\newcommand{\ee}{{\mathrm{e}}}

\journal{Journal of Luminescence}

\begin{document}

\begin{frontmatter}

\title{Biophotons, coherence and photocount statistics: a critical review}

\author[1]{Michal Cifra\corref{cor}}
\ead{cifra@ufe.cz}

\author[2]{Christian Brouder}
\author[1,3]{Michaela Nerudov\'{a}}
\author[1]{Ond\v{r}ej Ku\v{c}era}

\address[1]{Institute of Photonics and Electronics, Academy of Sciences of the Czech Republic, Prague, Czechia}
\address[2]{Institut de Min\'{e}ralogie, de Physique des Mat\'{e}riaux et de Cosmochimie, Universit\'{e} Pierre et Marie Curie Paris~6, CNRS UMR7590, Paris, France}
\address[3]{Department of Circuit Theory, Faculty of Electrical Engineering, Czech Technical University in Prague, Prague, Czechia}

\cortext[cor]{Corresponding author: Michal Cifra}

\begin{abstract}
Biological samples continuously emit ultra-weak photon emission (UPE, or ``biophotons'') which stems from electronic excited states generated chemically during oxidative metabolism and stress. Thus, UPE can potentially serve as a method for non-invasive diagnostics of oxidative processes or, if discovered, also of other processes capable of electron excitation. While the fundamental generating mechanisms of UPE are fairly elucidated together with their approximate ranges of intensities and spectra, statistical properties of UPE is still a highly challenging topic. Here we review claims about nontrivial statistical properties of UPE, such as coherence and squeezed states of light. After introduction to the necessary theory, we categorize the experimental works of all authors to those with solid, conventional interpretation and those with unconventional and even speculative interpretation. The conclusion of our review is twofold; while the phenomenon of UPE from biological systems can be considered experimentally well established, no reliable evidence for the coherence or nonclassicality of UPE was actually achieved up to now. Furthermore, we propose perspective avenues in the research of statistical properties of biological UPE.
\end{abstract}


\end{frontmatter}

\section{Introduction}

Ultra-weak photon emission (UPE, or ``biophotons'') from biological systems is a luminescent phenomenon which is present without any direct external stimulation nor additionally applied external luminophores \citep{cifra2014-jppb}. While there is some consensus about intensity and spectrum of UPE \cite{cifra2014-jppb,pospisil2014}, claims about statistical properties of UPE are very controversial. We aim to explain and settle this controversy in this critical review. %UPE can be viewed upon as a diagnostic method in agriculture \cite{kato2014} and biomedicine \cite{ou-yang2014} or as a phenonemenon which was proposed to mediate biocommunication \cite{prasad2014, cifra4, scholkmann2013}.  

Electronic excited states giving rise to UPE are generated chemically in the course of oxidative metabolic and stress processes \citep{cifra2014-jppb} in biological samples and living organisms. Several other terms synonymous to ultra-weak photon emission occur in the literature: autoluminescence \citep{havaux2006}, weak luminescence \citep{quickenden1974}, low level chemiluminescence \citep{cadenas2}, biophotons/biophoton emission \citep{cohen1997,devaraj1997}, \emph{etc.}
 Spectral range of UPE is known to lie at least in the range from 350 nm to 700 nm \citep{pospisil2014} and its intensity being up to several hundreds to thousand photons per square centimeter per second %cm$^{-2}$ s$^{-1}$ 
in the whole mentioned spectral range\footnote{The intensity of UPE in visible region of the spectrum is many orders of magnitude higher than the intensity of thermal radiation (described by Planck's law) for other parameters (sample area, temperature, \emph{etc.}) being the same, see \cite[Fig. 2]{cifra2014-jppb}.}.

From 1980s, there have been many claims about nontrivial statistical properties of UPE, such as coherence and even squeezed states of light \citep{Popp-86,Popp-94,Popp-09,Popp-03-Indian,Popp-02-non-classical,Bajpai-98,Bajpai-00}. Such properties of UPE would be of great physical and biological importance. At first, if the claims of UPE coherence were proved to be true, a novel mechanism of chemically powered ultra low power lasing would be very likely discovered. At second, there would be also great implications in biology since coherence or squeezed states of UPE would bring an evolutionary advantage for organisms in terms of ultra fast optical communication \citep{kucera2013} for purpose of intracellular and intercellular interactions and organization \citep{pokorny1}. 

Optical biocommunication has been targeted by several reviews \citep{prasad2014,scholkmann2013,kucera2013,cifra4,trushin2004,nikolaev2000}. 
Intensity and spectral properties of UPE have been also recently reviewed \citep{cifra2014-jppb,pospisil2014}. However, there is no critical review which covers detailed technical aspects of statistical properties of photon emission from biological systems.
Here we present development and current state of the literature on the statistical properties of UPE, especially focused on coherent and squeezed states of light and provide critical reflection of these works.% We show that above mentioned claims about coherent and non-classical properties of UPE have, in fact, poor foundations which are based on misinterpretation of photo-count statistics.

We first start with the Section \ref{sec-theory} to  present necessary theory of photocount measurement coherence and quantum optics and then we review the models used to analyze experimental distributions of UPE photocount statistics. 
In the Section \ref{sec-experiments}, we review the experiments of statistical properties of UPE and assess them from the point of view of current understanding of physics and biophysics. 
We found that although there are quite numerous papers which contain unsubstantiated claims about statistical properties of UPE, several high quality works can be also found. Based on reliable works, we propose future avenues in the research of the statistical properties of biological UPE in conclusion.

%This theoretical section \ref{sec-theory} explains basics may be skipped by quantum optics specialist but is essential for newcomers and non-specialists.

\section{Statistical properties of light}
\label{sec-theory}

\subsection{Theory of photocount measurement}
The statistical properties of UPE were mostly investigated experimentally by measuring the distribution of counts produced by UPE in a photodetector. Therefore, we briefly introduce the classical and quantum approach to photocount distributions ({\it photocount statistics} is another term often used in the literature).

\subsubsection{Classical theory}
The intensity of the light field averaged over a cycle of the oscillation is given by the expression~\citep[p.~86]{Loudon-00}
\begin{eqnarray}
\bar {I}(t) &=&  \frac{1}{2}\epsilon_0 c |E(t)|^2,
\end{eqnarray}
where $\bar {I}(t)$ is an intensity (irradiance) averaged over a cycle of oscillation with units W/m$^2$, $\epsilon_0$ is a permittivity of the vacuum, $c$ is velocity of light in the vacuum and $E(t)$ is an intensity of the electric field. Intensity can also be obtained as the time average of the Poynting vector perpendicular to the surface of the detector.

Let the efficiency of the detector be denoted by $\eta$. According to the semi-classical theory of
optical detection~\cite[p.~120]{Loudon-00}, there is a probability distribution $P(W)$ such that
the probability $p_n(t,T)$ of detecting $n$ photoelectric emissions in a finite time interval from
$t$ to $t+T$ is
 \begin{eqnarray}
p_n(t,T) &=& \frac{1}{n!} \left\langle W^n \ee^{-W}\right\rangle
  = \frac{1}{n!} \int_0^\infty W^n \ee^{-W} P(W) d W,
\label{Pntclass}
\end{eqnarray}
where $W=\eta \int_t^{t+T} I(\tau) d \tau$ and is integrated light intensity and $\eta$ is a coefficient containing dimensional factors
and describing the efficiency of the detector, so that $W$ is dimensionless.

\subsubsection{Quantum theory}
The quantum expression for the probability that $n$ photocounts occur between time $t$ and $t+T$ 
is~(\cite{Kelley-64}, \cite[p.~276]{Loudon-00}, \cite[p.~725]{MandelWolf})
\begin{eqnarray}
p_n(t,T) &=& \Big\langle {:}\frac{\hat{W}^n}{n!}
    \ee^{-\hat{W}}{:}\Big\rangle,
\label{Pntquant}
\end{eqnarray}
where 
\begin{eqnarray}
\hat{W} &=& \eta \epsilon_0 c 
\int_t^{t+T} 
  |\hat{E}(t)|^2,
\end{eqnarray}
in the Heisenberg representation.

All phenomena are basically of a quantum nature, but we say that a distribution of photocounts is classical if there exists a classical density distribution (\emph{i.e.} a non-negative $P(W)$) such that the (quantum) probability given by Eq.~(\ref{Pntquant}) is equal to the classical one given by Eq.~(\ref{Pntclass}). The characterization of non-classical light was investigated in detail~\citep{MandelWolf,Picinbono-05}. A probability of photocount detection is purely quantum if no such $P(W)$ exists.

\subsubsection{Conditional probability}
Some experiments are carried out with two photomultipliers (\cite[p. 87]{Popp-98}, \cite[chapter by X. Shen, p. 287]{bpe-book2003}). When a photon is detected by photomultiplier 1, the photons in photomultiplier 2 are registered during the time interval $\Delta t$. Bayes' theorem tells us that the conditional probability for $A$ given $B$ is given by $P(A|B)=P(A\cap B)/P(B)$.
The conditional probability of photocounts is calculated in \cite[p. 79]{Arecchi-69} and \citep{Ou-95}. The waiting-time distribution for coherent, squeezed and thermal states was investigated
in~\cite{Vyas-88}.

\subsection{States of the light and their photocount statistics} 
\subsubsection{Coherence and coherent states}
\label{coherence}
Coherence is a quite subtle property of light. In a nutshell, coherence is the ability of light to build interference which is, according to Grimaldi, the fact that darkness can be obtained by adding light to light\footnote{\emph{obscuratio, facta per solam additionem luminis}~\cite[p.~189]{Grimaldi}. In fact, Grimaldi did not really observe interference~\citep[p.~135]{Kipnis}, but his happy turn of phrase was remembered.}. 
Broadly speaking, light beams are coherent if they combine like waves (by adding the amplitudes of the beams) while they are incoherent if they combine like particles (by adding the intensities, \emph{i.e.} the square of the amplitudes,  of the beams).

It took a long time to clarify the meaning of coherence~\citep{Kipnis,Brosseau-09} and the coherence properties of even the most classical double-slit interference experiment are still a matter of active current research~\citep{Visser-08,Yadav-06,Agarwal-05}.
For example, the influence of coherence on the interference of light beams (\emph{i.e.} the Fresnel-Arago laws) was fully understood only in 2004~\cite{Mujat-04} and the conditions for a light beam to be considered as a sum of a fully polarized and a fully unpolarized beams are still controversial~\citep{Tervo-09}\footnote{The last example must be considered with a grain of salt because the concepts of polarization and coherence are not
identical~\citep{Jones-79}.}. The coherence of a light beam is modified by its propagation, the degree of coherence of a beam can influence its spectrum (this is known as the Wolf effect).

As a consequence of all these subtle effects, the literature on coherence is often very cautious and each statement is carefully supported by solid proofs. Many papers from the period 1980 - 2010 which aimed to study coherence of UPE often contain highly speculative statements.
Therefore, one of the main purposes of the current work is to assess the solidity of the conclusions drawn by the authors based on the data they presented and on the currently accepted physical viewpoints. We would also like to warn readers, especially those outside of the field of statistical properties of electromagnetic field, that some authors of original UPE literature use the term ``coherence'' rather vaguely. Terminology from quantum mechanics and quantum field theory often occurs in UPE literature, where the term coherence may refer either to (i) wave function from Schr\"{o}dinger equation or to (ii) light, which is, strictly speaking, not the same (although related) and often creates confusion. In quantum mechanics, coherence is an intrinsic property of the wave function and once decoherence occurs (\emph{i.e.} loss of wave function coherence or collapse of wave function), system behaves classically. Therefore, quantum behaviour is considered synonymous to coherence by some authors, but this is reasonable only when speaking about wave function.
We refer to coherence of light in this paper and one cannot directly equate terms \emph{non-classical (quantum)} to \emph{coherent} light, neither \emph{classical} to \emph{incoherent} light. Generally, quantum optical framework can explain all states of light. Classical framework can explain only some of them and those can be called classical. The states which can be explained in quantum framework only are usually called quantum states. Coherence of light can be both of classical and quantum character, thermal states of light (see further down) can be described in classical and quantum framework, while certain states can be described only in quantum framework (\emph{e.g.} squeezed states).

\subsubsection{Classical coherence}
If $\bfE_1(\bfr,t)$ and $\bfE_2(\bfr,t)$ are the electric fields of two light waves (\emph{i.e.} two solutions of the
Maxwell equations), then the linearity of the Maxwell equations implies that the two waves add to form a new light wave: $\bfE(\bfr,t)=\bfE_1(\bfr,t)+\bfE_2(\bfr,t)$. In particular, if $\bfE_1(\bfr,t)=-\bfE_2(\bfr,t)$, then $\bfE(\bfr,t)=0$. In other words, as Gabor put it when he received his Nobel prize in physics for the invention of holography: ``light added to light can produce darkness''. It might not be completely out of place to recall that Gabor and Reiter devoted a book on the radiations emitted by plants and their influence
on cell division, where they observed diffraction~\cite[p.~20]{Gabor-28} and reflection~\cite[p.~21]{Gabor-28} of UPE (but they did not investigate coherence).
In 1928, they stated: ``die Existenz der Strahlung bestimmter biologischer Objekte und die Wirkung dieser Strahlung auf die Zellteilung steht nach unseren Versuchen au{\ss}er allem Zweifel.'' (translation: \emph{the existence of the radiation of certain biological objects and the effect of the radiation on the cell division is beyond any doubts according to our experiments.})~\cite[p.~6]{Gabor-28}.
Much later (in 1956),  he wrote: ``The results \dots seemed to support\dots the hypothesis of some radiating
agency; on the other hand all experiments for proving the radiation by physical means have failed.
To this day (1956) nobody knows what these experiments really mean.''~\citep{Allibone-80}.

Light detectors do not resolve the time-dependence of electromagnetic fields and measure something
which is proportional to an average of the intensity over a duration $\Delta t$: $I(\bfr,t)=(1/\Delta t) \int_t^{t+\Delta t}|\bfE(\bfr,\tau)|^2 d\tau$. 
The light beam is coherent if the total intensity of the two beams is obtained by adding the amplitudes:
$I(\bfr,t)=(1/\Delta t) \int_t^{t+\Delta t} |\bfE_1(\bfr,\tau)+\bfE_2(\bfr,\tau)|^2 d\tau$. 
It is incoherent if the total intensity of the two beams is obtained by adding the intensities
$I(\bfr,t)=(1/\Delta t) \int_t^{t+\Delta t} |\bfE_1(\bfr,\tau)|^2+|\bfE_2(\bfr,\tau)|^2 d\tau
=I_1(\bfr,t)+I_2(\bfr,t)$. 
The reason why intensities should be added is not entirely clear in classical optics\footnote{As noticed by Biedenharn and Louck for the related problem of unpolarized radiation: ``Every solution of the Maxwell equations, which propagates spatially as a plane wave, is necessarily completely polarized transversally; every additive superposition of two completely polarized solutions yields another completely polarized solution. An unpolarized wave cannot be a solution of the Maxwell equations! Thus, the concept of an unpolarized wave goes beyond Maxwell electrodynamics and involves quantal considerations.'' \cite[p. 453]{BL}}.

Coherence is not a yes-or-no attribute but a continuum-like. Strictly speaking, any electromagnetic (light) field is coherent to certain extent. Coherence time $T_c$ or coherence length $L_c$ is often used to describe the extent of coherence \cite[sec.7.5.8]{bornWolf1985}. Within the $T_c$, the time dependence of any light field at a point in space can be very closely approximated by a sine wave, \emph{i.e.} the field is coherent. In practice, we say that the light is coherent if it displays very large coherence time (\emph{i.e.} much larger than the period of the oscillation) or very large coherence length (\emph{i.e.} much larger than the wavelength), such that interference effects can be observed. Relation between coherence length and coherence time is $L_c =\textrm{c} T_c ~\approx \textrm{c}/\Delta f = \lambda_0^2/\Delta\lambda$, where c is the velocity of light, $\lambda_0$ is the mean wavelength,  $\Delta f$ and $\Delta\lambda$ is the spectral bandwidth of the light in Hz and nm, respectively. The broader the spectral range emitted from the source, the shorter the coherence time and coherence length: $T_c \cdot \Delta f \geq 1/4\pi$ \cite[p.~319]{bornWolf1985}.

\subsubsection{Quantum coherence}
We follow the discussion of coherence described by Mandel \cite[ch.~12]{MandelWolf}.
If $\mathbf{\hat{E}^{(+)}}(\mathbf{r},t)$ and $\mathbf{\hat{E}^{(-)}}(\mathbf{r},t)$ is the annihilation and creation operator of the electric field, respectively~\citep[p.~574]{MandelWolf}, then the intensity of light at a point is proportional to
$\langle \mathbf{\hat{E}^{(-)}}(\mathbf{r},t) \mathbf{\hat{E}^{(+)}}(\mathbf{r},t)\rangle$, where the sign $\langle \cdot\rangle$ represent the expectation value over a quantum state, which can be a mixed state.
More generally, a quantity such as 

%$\langle \mathbf{\hat{E}^{(-)}}(\mathbf{r}_1,t_1)\dots \mathbf{\hat{E}^{(-)}}(\mathbf{r}_k,t_k) \mathbf{\hat{E}^{(+)}}(\mathbf{r}'_1,t'_1)\dots \mathbf{\hat{E}^{(+)}}(\mathbf{r}'_l,t'_l) \rangle$ 

$\langle \mathbf{\hat{E}^{(-)}}(\mathbf{r}_1,t_1)\dots \mathbf{\hat{E}^{(-)}}(\mathbf{r}_N,t_N) \mathbf{\hat{E}^{(+)}}(\mathbf{r}'_M,t'_M)\dots \mathbf{\hat{E}^{(+)}}(\mathbf{r}'_1,t'_1) \rangle$ 
is called a \emph{correlation function}, \cite[p.~585]{MandelWolf}.
A state for which there is a vector function $\bfe(\bfr,t)$ such that 
\begin{eqnarray*}
\langle \mathbf{\hat{E}^{(-)}}(\mathbf{r}_1,t_1)\dots \mathbf{\hat{E}^{(-)}}(\mathbf{r}_N,t_N) \mathbf{\hat{E}^{(+)}}(\mathbf{r}'_M,t'_M)\dots \mathbf{\hat{E}^{(+)}}(\mathbf{r}'_1,t'_1) \rangle &=&
\end{eqnarray*}

\begin{eqnarray*}
 \mathbf{e}^*(\mathbf{r}_1,t_1)\dots \mathbf{e}^*(\mathbf{r}_N,t_N) \mathbf{e}(\mathbf{r}'_M,t'_M)\dots \mathbf{e}(\mathbf{r}'_1,t'_1) \ & &
\end{eqnarray*}
is said to \emph{factorize}.

Such a state corresponds to full coherence in the classical case. As we shall see, all correlation functions
factorize if the system is in a \emph{coherent state}. The reciprocal question (\emph{i.e.} the determination of all the states that lead to factorized correlation functions) was studied by Honneger and Rieckers~\cite{Honegger-01,Honegger-04}. They found out that there are states for which all
correlation functions factorize and which are not coherent states in the usual sense. Very general coherent states were defined in the mathematical literature~\citep{Skoda-07}, but they were not used yet in the present context.

\subsubsection{Coherent states}
Coherent states were discovered by Schr\"{o}dinger \cite{Schrodinger-26-coherent}, rediscovered by Schwinger \cite{Schwinger-53}, then called \emph{coherent states} and further studied by Glauber \cite{Glauber-63}. Coherent states are now a standard tool of quantum optics~\citep{MandelWolf}.
From a mathematical point of view, a coherent state $|\alpha\rangle$ is an eigenstate of the annihilation operator: $\hat{a} |\alpha\rangle=\alpha|\alpha\rangle$, where $\alpha$ is a complex number~\cite[p.~523]{MandelWolf}. From the conceptual point of view, coherent states are the quantum states that correspond to classical electromagnetic waves. For instance, a classical current (a piece of electric wire carrying a macroscopic current, for instance) gives rise to a coherent state of the photon field~\citep{Itzykson}.

\begin{figure}
\begin{center}
\includegraphics{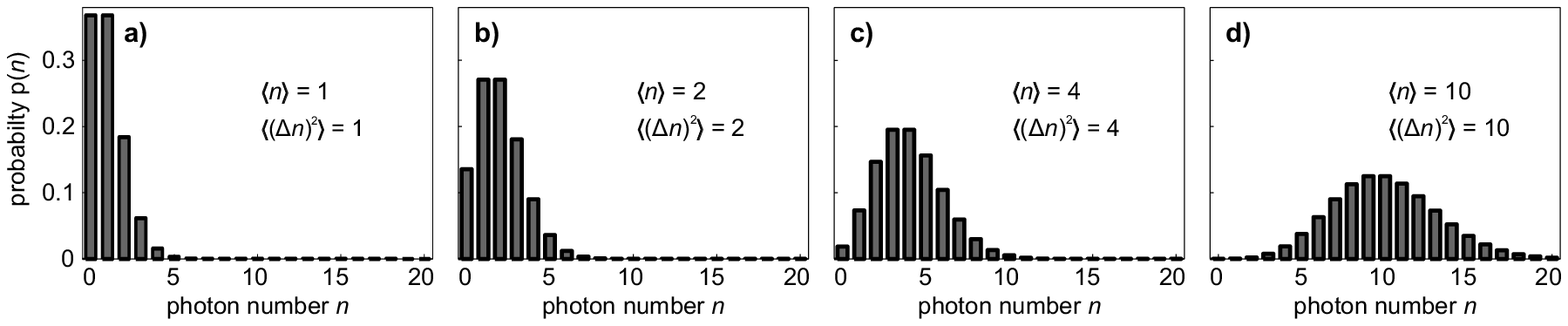}
\caption{Poisson photocount distribution is assymetric for low intensities of light flux. Here we show Poisson distribution for four different average values of photon counts $\langle n\rangle$. {\bf Formatting note:  Image is out of the printing area because of the submission formatting of the manuscript. See all images after the text of the manuscript for the full image} \label{figpoisson}}
\end{center}
\end{figure}

The photocount statistics of a system in a coherent state is a Poisson distribution (see Fig.~\ref{figpoisson})
\begin{eqnarray*}
p_n(t,T) &=& \frac{\langle n\rangle^n}{n!} e^{-\langle n\rangle},
\end{eqnarray*}
where $\langle n\rangle=T\langle \dot{n}\rangle$ is the average number of photons measured between time $t$ and time $t+T$. A Poisson distribution is a sign of classical light field. Its variance is equal to its mean: $\langle (\Delta n)^2 \rangle=\langle n \rangle$. The departure from a Poisson distribution is measured by the Fano factor $F$ such that $\langle (\Delta n)^2 \rangle=\langle n \rangle F$ or by the Mandel parameter $Q=F-1$.
A photocount statistics is super-Poissonian if $F>1$ and $Q>0$, it is sub-Poissonian (and therefore non-classical) if $F<1$ and $Q<0$. Departure from Poisson distribution is a sign of non-classical (quantum) nature of light.

Note that a laser light is not in a coherent state~\citep{Pegg-05}, although it has a very large coherence length and a pronounced phase coherence~\citep{Pegg-09}. Moreover, its probability distribution
can be far from Poissonian~\cite[p.~940]{MandelWolf}.

\subsubsection{Squeezed states}
\begin{figure}
\begin{center}
\includegraphics{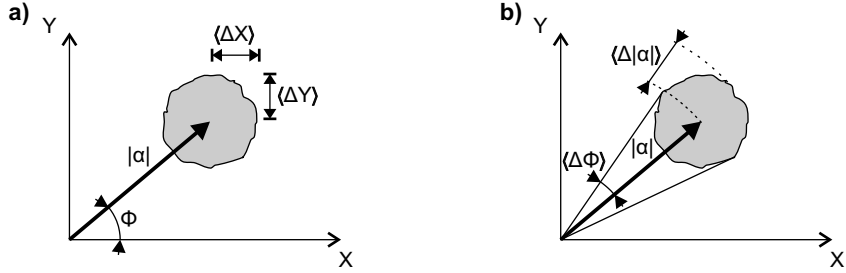}
\caption{Uncertainty region (in grey color) in phase space. Uncertainty region actually corresponds to the contour of Wigner function given by ca. 1 standard deviation from its center. \textbf{a)} Uncertainity of the canonical variables, here quadrature components X and Y, is expressed as their standard deviations $\langle\Delta X\rangle$ and $\langle\Delta Y\rangle$, respectively. \textbf{b)} Uncertainties in the phase and amplitude of the field can be expressed through their standard deviations $\langle\Delta \phi\rangle$ and $\langle\Delta \alpha\rangle$, respectively. For the measurement of the photocount statistics, it is useful to note that the photon number standard deviation (\emph{i.e.} standard deviation of the photocount distribution) is related to the standard deviation of a field amplitude  approximately as $\langle \Delta n\rangle \approx 2 \langle n\rangle^{1/2}  \langle \Delta |\alpha| \rangle$ \citep[eq.~9]{teich1989}.}. \label{fig-uncertainty}
\end{center}
\end{figure}

Squeezed states have the characteristic that the dispersion (uncertainty) of one variable is reduced at the cost of an increase in the dispersion of the other canonical variable, for instance position vs. momentum or amplitude vs. phase\footnote{It turns out that it is impossible to generate the phase operator which would be Hermitian \cite[p.~492]{MandelWolf}. Note that the property of being ``Hermitian'' is necessary for the operators used in quantum mechanical calculations. Instead of phase operator, cosine and sine operators, which can be generated as Hermitian, are used to work with the phase properties of the field.}. Following Loudon \cite{loudon1987}, this can be easily visualized when we write equation for electric field operator of a single mode of the photon field as
\begin{equation}
{\bf \hat{E}}(\bfr,t) = \bfE_0 \left[\hat{X} \sin(\omega t - \bfk \bfr) - \hat{Y} \cos(\omega t - \bfk \bfr)\right]
\end{equation}
where $\bfE_0 $ is amplitude of vectorial electric field, $\hat{X}$ and $\hat{Y}$ are Hermitian operators related to annihilation and creation operators of the photon field as $\hat{X} = (\hat{a} + \hat{a}^\dagger)/2$ and $\hat{Y}=(\hat{a}-\hat{a}^\dagger)/2\mathrm{i}$. $\hat{X}$ and $\hat{Y}$, the real and imaginary parts of the complex amplitude, give dimensionless amplitudes for the two quadrature phases. They obey the commutation relation $[\hat{X},\hat{Y}] = \mathrm{i}/2$. Electric field can be then depicted in a complex plane, see Fig. \ref{fig-uncertainty}. 

Various squeezed states were used in the UPE literature, but the most general ones are called two-photon coherent states and were proposed in 1976 by Yuen \cite{Yuen-76}. They have become standard states of quantum optics~\cite[p.~1046]{MandelWolf}. They are simply defined as the solution $|\alpha,\xi \rangle$ of eigenvalue equation $\hat{A}|\alpha,\xi \rangle = \beta|\alpha,\xi \rangle$. We follow the notation from Orszag \cite[ch.~5]{orszag2008}:

\begin{eqnarray}
\hat{A} &=& \mu \hat{a} + \nu \hat{a}^\dagger\\ 
\beta &=& \mu \alpha +\nu \alpha ^\ast \\
\nu &=& \ee^{\mathrm{i} \theta} \sinh r \\
\mu &=& \cosh r\\
\alpha &=& |\alpha| \ee^{\mathrm{i}\phi}\\
\xi &=& r \ee^{\mathrm{i}\theta}
\end{eqnarray}

\begin{figure}
\begin{center}
\includegraphics{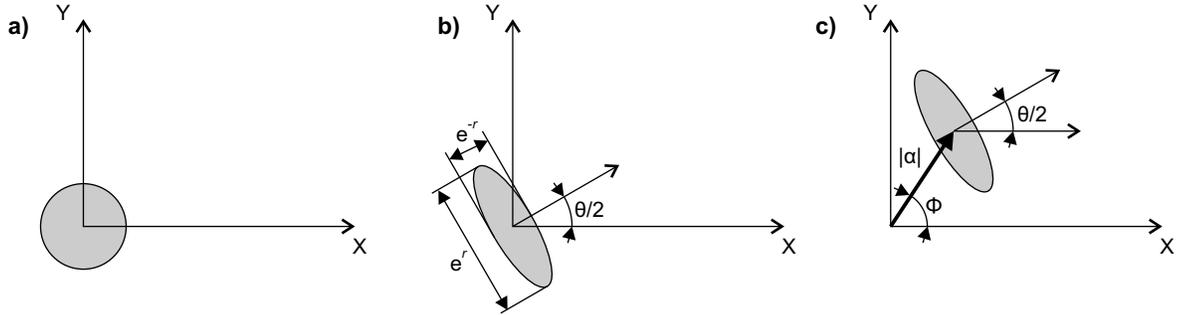}
\caption{Generation of a coherent squeezed state $| \alpha, \xi\rangle$ from a vacuum state $|0\rangle$. \textbf{a)} Region of uncertainity of the canonical variables of vacuum state is represented by a circle at the origin of the phase space. \textbf{b)} Application of the squeezing operator on the vacuum state leads to squeezing of the uncertainty region into the ellipse and the rotation of it by $\theta/2$. \textbf{c)} Application of the displacement operator simply shifts the uncertainty region by $|\alpha|$ in the direction given by the angle $\phi$.}. \label{fig-squeezing}
\end{center}
\end{figure}

Squeezed states, as originally defined by Yuen \cite{Yuen-76}, were produced by squeezing the coherent state. It means that at first, displacement operator $\hat{D}(\alpha)$ is used to create coherent state $|\alpha\rangle$ from vacuum state ($|0\rangle$) and then the squeezing operator $\hat{S}(\xi)$ is used, see \cite[Fig 21.3.b, p.1043]{MandelWolf}. Mathematically: $\hat{S}(\xi)\hat{D}(\alpha)|0\rangle = \hat{S}(\xi)|\alpha\rangle = |\xi, \alpha\rangle$. However, it is more convenient, and also often used in modern literature, to apply first the squeezing operator on the vacuum state and then the displacement operator: $\hat{D}(\alpha) \hat{S}(\xi)|0\rangle = \hat{D} (\alpha)|\xi\rangle = | \alpha, \xi\rangle$. Such procedure which gives so called ideal squeezed state (see \cite[Fig.~21.3.a, p.1043]{MandelWolf} and Fig.~\ref{fig-squeezing} in this paper), was introduced in reference \citep{caves1981} and we use it also in this paper.

The photocount statistics of a light field in a coherent squeezed state is given in terms of Hermite polynomials~\cite[p.~21]{walls2008}

\begin{eqnarray}
p_n(t,T) &=& {\left( n!\cosh r\right)^{-1}}
 {\left[\frac{1}{2}\tanh r \right]^n} 
\ee^{\left(-|\alpha ^2 |- \frac{1}{2} \tanh r \left(\left( \alpha ^ \ast \right)^2 
\ee^{\mathrm{i} \theta} + \alpha ^2 \ee^{-\mathrm{i} \theta} \right)\right)}
{|H_n \left(z \right)|^2} ,
\end{eqnarray}
\textrm where
\begin{eqnarray*}
z &=& \frac{\alpha + \alpha ^ \ast \ee^{\mathrm{i} \theta} \tanh r}{\sqrt{2 \ee^{\mathrm{i} \theta} \tanh r}} ,
\end{eqnarray*}

\begin{figure}
\begin{center}
\includegraphics{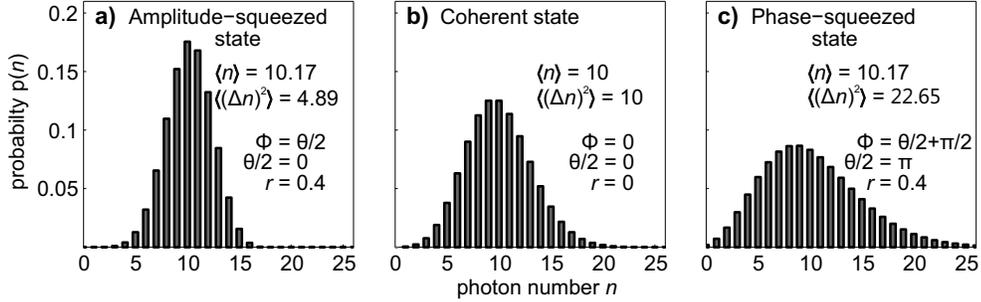}
\caption{Shape of the photocount distribution of the squeezed state depends on the squeezing parameters $r$ and rotation angles $\phi, \theta$, see their meaning explained in Fig. \ref{fig-squeezing}. Here we show quadrature-squeezed coherent state photocount distribution for \textbf{a)} Amplitude-squeezed light, where fluctuations of amplitude (\emph{i.e.} number of photon counts) are reduced - the distribution is narrower than that of Poisson distribution, \textbf{b)} coherent light (Poisson distribution) for comparison, \textbf{c)} phase-squeezed light, where phase fluctations are reduced (not possible to depict in photocount distribution), which inevitably leads to increase of amplitude fluctuations - the photocount distribution is broader. \label{figsqueezed}}
\end{center}
\end{figure}

Squeezed states are interesting because they can manifest lower intrinsic noise (fluctuations around mean of a canonical variable) than coherent light \citep{yuen2004}, a feature which classical light with Poisson distribution cannot achieve. See Fig. \ref{fig-squeezing} for manifestation of squeezing in the photocount distributions. The lower the intrinsic noise (related to uncertainty of a canonical variable), the higher the capability of such states to transmit information \citep{saleh1987}, but there are natural limits \citep{yuen2004}. However, squeezed states are fragile. They can be ``destroyed''  by interaction with environment, such as attenuation, beam splitter or a mirror, as those admit the vacuum fluctuations, which exceed the squeezed fluctuations, to enter from outside.

\begin{figure}
\begin{center}
\includegraphics{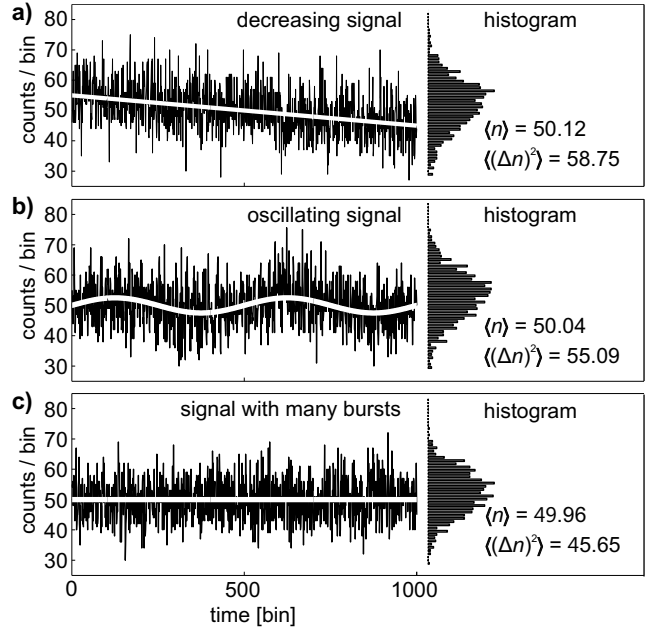}
\caption{Non-stationary signals can generate super- and sub-Poisson distributions from Poissonian signals. Random signals with Poisson distribution have been generated by computer (Matlab, function {\it poissrnd}) and combined with other signals to simulate specific experimental artifacts. \textbf{a)} Poisson signal superposed to a decreasing trend, \textbf{b)} Poisson signal superposed to an oscillating trend, \textbf{c)} Poisson signal with 200 random (uniform distribution) bursts with values from the interval from 20 to 40. Bursts were placed in random positions in the signal and their values substituted the original values at these positions}. 
 \label{fignonstationarity}
\end{center}
\end{figure}

One has to be careful to avoid experimental and instrumental artifacts before interpreting the photocount statistics data. Non-Poisson distribution of photocounts can be also generated by classical and thermal light which would otherwise lead to Poisson distribution if the measurement has been performed correctly. Trivial manifestation of non-Poisson distribution can be caused by non-stationarity of the light source such as modulation of the intensity of the photon signal due to the (i) slow drifting (Fig.~\ref{fignonstationarity} a) or periodic (Fig.~\ref{fignonstationarity} b) trends, (ii) random small bursts caused by electronic noise or by photon emission caused by other intensity limited stochastic processes (Fig.~\ref{fignonstationarity} c) and added to the Poisson signal, (iii) thresholding the pulses from the photodetector, \emph{etc.} Non-Poisson distribution caused by such non-stationarities and photon signal deformation has nothing to do with squeezed state of light.

\subsubsection{Thermal states}
A thermal state of light can be physically obtained by filtering thermal radiation. The photocount statistics of a thermal source with $M$ modes (degrees of freedom) is well approximated by the expression~\citep[p.~680]{MandelWolf}:
\begin{eqnarray}
p_n(t,T,M) &=& \frac{(n+M-1)!}{n!(M-1)!}
   \Big(1+\frac{M}{\langle n\rangle}\Big)^{-n}
   \Big(1+\frac{\langle n\rangle}{M}\Big)^{-M},
\end{eqnarray}
where $\langle n\rangle$ is the average number of photons and $M$ is the number of field modes (see also \cite[p.~731]{MandelWolf}). The number of degrees of freedom $M$ can be estimated by the product of a time degeneracy $M_t$ and a space degeneracy $M_s$. The time degeneracy is the ratio of the measurement time (bin time) over the coherence time~\cite[p.~97]{Loudon-00}. Thermal states are classical. An important characteristic of these states is the relation between variance and mean:

\begin{eqnarray}
\label{eq:varmeanMultimode}
\langle (\Delta n)^2\rangle &=&
  \langle n \rangle + \frac{\langle n\rangle^2}{M}.
\end{eqnarray}

The coefficient $M$ is generally very large for chaotic sources~\citep{Jiang-03-CCD}, so that the relation between variance and mean is close to that of a coherent state. Another interpretation is to say that, for reasonable intensities, the expectation value of the number operator $n$ is very small and the expectation value of $n^2$ is therefore negligible with respect to that of $n$~\cite[p.~19]{Mosset-04-PhD}. As a matter of fact, for large $M$, $p_n(t,T,M)$ tends to a Poisson distribution of parameter $\langle n\rangle$. 

Since the question whether field giving rise to UPE is in a coherent or a thermal state is recurrent in the UPE literature, it is important to know whether photocount statistics can distinguish between them. Since photocount statistics of thermal light becomes equal to that of a coherent state when number of modes M is large, photocount statistics is not able to discriminate between a coherent and a thermal state with many modes. This can be seen in particular in the relation between variance and mean in a thermal state, see Eq.~\ref{eq:varmeanMultimode}.

\begin{figure}
\begin{center}
\includegraphics{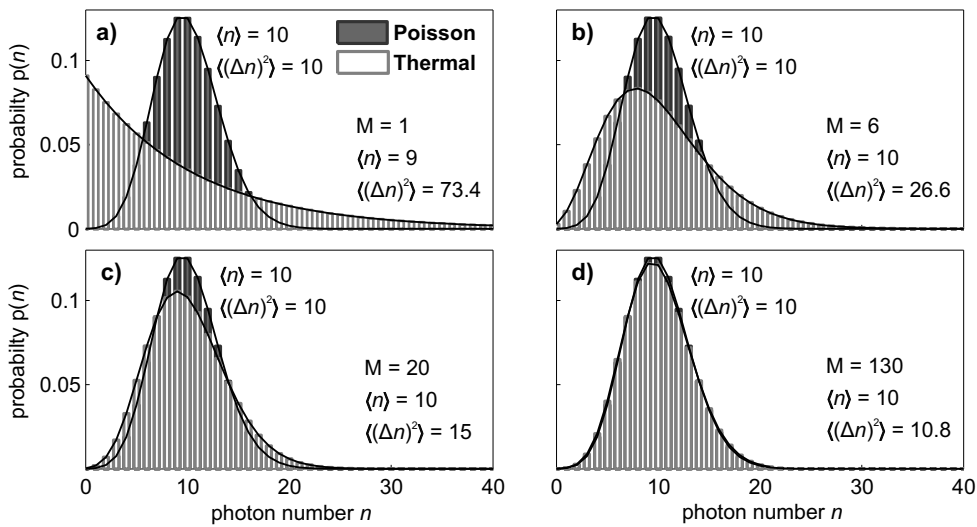}
\caption{Thermal field photocount distribution (light grey) approaches Poisson distribution (dark grey) for large number of modes M, \emph{i.e.} for large number of independently radiating sources (molecules, atoms) or from single source with very short coherence time compared to time interval of measurement. Average value of intensity of the photon signal $\langle n \rangle$ is the same for all displayed distributions.
  \label{figthermal}}
\end{center}
\end{figure}

\subsubsection{Superposition of coherent and thermal states}
If UPE contained mixture of coherent and thermal states, corresponding description of such superposed fields should be used. This has not been done in UPE literature up to our knowledge, although photocount statistics of superpositions of coherent and thermal states was already investigated by Perina \cite{Perina-67,Perina-10}. 

\subsubsection{Super-radiance}
\label{superrad}
Super-radiance is the coherent emission of light by several sources. It was first proposed by Dicke \cite{Dicke-54} and is now a thoroughly investigated subject~\citep{Hepp-73,gross1982,Benedict-96,Emary-03}.
Its main characteristic is the fact that the intensity of the emitted light can vary with the square of the number of sources because they can emit in phase.

The photocount statistics of super-radiant emission was investigated in detail by \cite{Trifonov-99}, who found cases where the statistics is sub-Poissonian (see also \cite{Benedict-96}, \cite{Benedict-96}, sections 1.3 and 11.6). \cite{Nagashima-78} observed that the photon state of a super-radiant system is generally not a coherent state\footnote{This contradicts a previous paper by Bonifacio \cite{Bonifacio-70}, who overlooked the contribution of non-diagonal terms.}.

\section{Experimental works on the photocount statistics of UPE}
\label{sec-experiments}
Photoelectric measurement of UPE were attempted already from the early 1930s (see \cite{Lorenz-34}, \cite{Gray-33}, \cite{Gray-33}, and references therein), but reliable measurements could be obtained only in the early 1950s~\citep{Strehler-51,colli1}. The very delicate instrumental aspects of UPE photocount measurements were discussed in several papers~\citep{Ruth-77,Kobayashi-00,Kobayashi-03,Kobayashi-03-conf,Shen-03,SchirmacherPhD}.

First, it is instructive to refer to several relevant works dealing with statistical properties of luminescence from non-biological sources. The photocount statistics of weak luminescent sources was measured for solid-state ZnS:Cu luminophores~\citep{Konak-82}, luminescent glass~\citep{Konak-82} and single molecules in microdroplets~\citep{Hill-98}. All these experiments can be analyzed in terms of thermal source or Poisson statistics. The photocount statistics of light emitting diodes was found to be Poissonian~\citep{Kobayashi-98} or super-Poissonian in case of avalanche photodiodes operated above its breakdown voltage and used as a light source~\citep{Huang-05}. Chemiluminescence of a 9,10-diphenylanthracene radical ions in acetonitrile solution shows Poisson statistics~\citep{Collinson-95}.

For the following discussion, it is important to stress that Poisson statistics is not a proof of the existence of a coherent state of light. According to the Palm-Khintchine theorem, the superposition of a large number of independent equilibrium renewal processes, each with a small intensity, behaves asymptotically like a Poisson process. For example this is true for the limit $X(t)=\sum_{i=1}^n x_i(nt)$, where the processes
$x_i$ are independent~\citep{Lindner-06}. However, this result depends on the way the limit is taken~\citep{Cateau-06,Lindner-06}. We suggest that superposition of random nonstationary emissions, which was investigated by \cite{Banys-77} and \cite{Saleh-83}, seems to be the most reasonable first approach for modeling biological chemiluminescence.

\subsection{Works with conventional data interpretation}
There are several works on the UPE photocount statistics that are at the qualitative level of the quantum optics literature, without over-interpretation of the results. Papers of Kobayashi and Inaba belong to this category. 
They performed \cite{Kobayashi-98} an interesting and useful investigation of the photocount statistics of a time-dependent system. The influence of cosmic rays and microdischarges is taken into account to get reliable data. For a light emitting diode they found a Fano factor close to one, for the photoluminescent bacterium \emph{Photobacterium phosphoreum} they measured a Fano factor significantly greater than one, indicating a super-Poissonian statistics. More precisely: ``During the primary stage of cell proliferation, the photon statistics show super-Poisson behavior, which changes to Poisson statistics according to the increase in the number of cells''. The Fano factor is analyzed in terms of a chaotic source (using Eq.~(41) of ref.~\cite{Saleh-83} and denotes a ``clustering of excitation and emission''.

In another paper \cite{Kobayashi-00}, they describe the experimental setup in great detail. The long-term stability of the dark counts is checked, as well as their power spectrum, their statistics as a function of counting time, their auto-correlation function and the dependence of it on the photomultiplier. The regenerative effects, cosmic rays and microdischarges are also taken into account. 
The authors measured the temperature-dependence of the dark counts and made a careful analysis of how
the dark counts should be subtracted. They compare the corrected $g^{(2)}(\tau)$ (second order correlation function) with experiment for randomized laser light of various intensities. They discuss the measurement of the Fano factor in the presence of dark current and for a time-dependent source. Finally, they measure the photon statistics of \emph{Dictyostelium discoideum}. Variation of the Fano factor during the early stage of development and after starvation is observed. Further, they found super-Poisson statistics (\emph{i.e.} photocount distribution with a width greater than a Poissonian distribution and Fano factor $>$ 1), which they interpreted, as in their previous work, to be caused by clustering of excitation and emission processes where the optical field is composed of a sequence of independent flashes initiated by Poisson random time events. No relation to squeezed states, which can also manifest super-Poisson statistics, was mentioned. This article \cite{Kobayashi-00} represents a quality benchmark for all UPE photocount measurements in terms of careful verification of the experimental setup and rigorous interpretation of the data.

Kobayashi \emph{et al.} also discussed the measurement of photocount statistics with 2D-photomultipliers~\citep{Kobayashi-99,Kobayashi-03,Kobayashi-05}. Note that Inaba \emph{et al.} measured UPE images already in 1988 (\cite{Scott-89}, see also refs.  \cite{Inaba-90}, \cite{Kobayashi-96}, \cite{Kobayashi-97}).

Another remarkable publication on this subject is the PhD thesis by Erich Schirmacher~\cite{SchirmacherPhD}. He made very careful experiments and a thorough theoretical analysis. He measured photon statistics from samples of lichen (\emph{Parmelia physodes}) covering a tree bark, a leaf from a dark plum tree (\emph{Prunus cerasifera} `Nigra'), leaves on a twig from silver fir (\emph{Abies alba}), a leaf from baynan tree \emph{Ficus microcarpa}, a leaf of a stinging nettle \emph{Urtica dioica} and a leaf from oak \emph{Quercus robur} that he compared to the light beam of a He-Ne laser. He observed only super-Poissonian statistics and did not find conclusive evidence of a non-classical (quantum) behavior of light.

There are several other works which provided UPE photocount statistics without speculative intepretation. Williams \emph{et al.} measured UPE from human breath and observed a photocount distribution with two peaks~\citep{Williams-82,Slawinska}. This interesting experiment should certainly be reproduced. This is not really UPE from a living organism, but this effect could create an artifact in the measurement of UPE from human beings. Shen \emph{et al.} \cite{Shen-93} measured the photocount statistics of cucumber seedlings, mungbean seedlings and rhizobium bacteroids. They conclude that: ``Experimental evidence accumulated so far leaves no doubt as to the validity of the biochemical interpretation of the chemi-excitation and its association with metabolism in biological systems.'' Similarly, Gallep measured many different samples and analyzed his results in a rational way~\citep{Gallep-04,Gallep-07}. Van Wijk \emph{et al.} made use of Bajpai's coherent states to fit experimental photocount statistics. This enabled them to distinguish UPE from various parts of a human body~\citep{Wijk-06,Wijk-10,bajpai2013}. These papers already contain unjustified speculations about squeezed states of UPE.

\subsection{Works with speculative data interpretation}
There are several researchers who pursued unconventional and speculative interpretation of the UPE experiments and photocount statistics, mainly based on the hypothesis of coherent processes in biological systems. We analyze the evolution of two main streams of ideas of coherent states and squeezed states of biological light chronologically. See Appendix A for the reference and selected comments on the works of other authors in the category of speculative data interpretation.

\subsubsection{Coherence of ultra-weak photon emission ?}
Work on statistical properties of biological ultra weak photon emission focused on coherence was pioneered by Fritz-Albert Popp. Activities of F.-A. Popp attracted many scientists and also public interest to the topic of biophotons. However, his interpretations experimental results on UPE photocount statistics in terms of coherent states are controversial and therefore are not generally accepted in scientific community.

Bernhard Ruth, supervised by Popp, built an efficient photomultiplier-based measurement system of UPE. Within his thesis~\citep{Ruth-77} he showed that many biological samples are source of ultra-weak photon emission. The most controversial result of this thesis is a series of  UPE spectra~\citep{Ruth-76,Ruth-77}, that are completely different from UPE spectra measured later~\citep{Hideg-90,Kobayashi-03}. Care needs to be taken because possible artifact leading to these strange spectra is the luminescence of the filters~\citep{Schmidt-87}.

In this period, the working hypothesis was introduced: the biological UPE originates from biological coherent photon field ~\citep{Popp-77} which regulates biological processes. This hypothesis was inspired mainly by following points:
\begin{itemize}
\item Several polycyclic hydrocarbons have been investigated. Correlations between their electronic properties and carcinogenic activity have been found \citep{Popp-76}. Popp proposed that the mechanism of the action of the cancerogenic substances is the disturbance of the excitation cellular photon field at certain energy which is related to DNA repair \cite{Popp-76}, \cite[p.~117]{bpe-book2000}.
\item Coherent electrically polar vibration states in GHz-THz region in metabolically active cell have been postulated by Fr\"{o}hlich \cite{froh6,froh2,froh9,froh10}. Popp embraced the general idea of coherent processes in biology and assumed based on the model of Li \cite[ch.~5]{bpe-book1992} that the DNA in cells behaves as a low level excimer laser generating coherent photon field.
\
\end{itemize}

From that  time on, experimental data obtained in Popp's group and their followers have been attempted to fit the coherence theory of biological ultra weak photon emission. In the Table \ref{table:popp}, we highlight several specific points from these works which are the most controversial and deviate most strongly from currently accepted knowledge in order to inform readers where the caution should be exercised.

%%%%%%%%%%%%%%%%%
\begin{longtable}[h]{|p{9cm} | p{3cm} |}
\caption{Chronologically ordered assessment of publications of F.A. Popp and the statements contained there which are the most controversial and deviate most strongly from currently accepted knowledge \label{table:popp}}\\

%\begin{table}[h]
%\begin{center}
%\begin{tabular}{|l|c|}
\hline
Statements, issues & Main references\\
\hline
UPE statistics of variously stressed cucumber seedlings was measured and analyzed in terms of chaotic light. In many cases the photocount distribution was far from Poissonian. The body of the paper brings plenty of data. The part of the conclusion involves statement which is not substantiated by the data in the paper. For instance, the authors state that DNA is the origin of UPE: ``DNA may represent active photon stores which are governed by Bose condensation''~\cite[p.~312]{Popp-81}. & \cite{Popp-81} \\ 
\hline
Further unfounded statements. For instance, often a statement is found that erythrocytes (red blood cells) do not emit UPE because they do not contain DNA. This argument is used to support the hypothesis of UPE generation by DNA. Although this statement is important no reference to the source of experimental data is found. It should be noted that erythrocytes have also many other differences in their structure compared to other cell types than the presence of nucleus. Only mammalian erythrocytes, compared to vertebrates, do not contain nuclei as well as other organelles such as mitochondria, Golgi apparatus and endoplasmic reticulum. & \cite{Popp-81-wieder,Popp-83-Arzt,Popp-84,Popp-86,Popp-86-Pasteur,Popp-95}\\
\hline
``Measurements of photo count statistics show that the probability of registering $n$ photons within a given time interval $\Delta t$ is significantly different from a purely chaotic distribution, even for a multimode system with the highest possible degree of freedom.''~\cite[p.~119]{Li-83} Although they do not show any comparison with experiment, they add: ``On the other hand, the consistency of the results with a Poisson distribution, which accounts for a coherent radiation field, cannot be refused,''~\cite[p.~119]{Li-83}. As we saw, a Poisson distribution is indeed compatible with a coherent state, but also similar to other states of light. & \cite{Li-83} \\
\hline
Papers where authors were trying to prove that the hyperbolic decay of delayed luminescence is a ``sufficient condition for coherence''. They start from a harmonic oscillator ~\citep{Li-83,Popp-84}. There have been several conceptual and mathematical mistakes identified in these papers, see ref.~\citep{Salari-11} for a detailed investigation of one of these papers. Further, it needs to be stressed that the state of stationary UPE (autoluminescence), where light is generated by some biochemical reaction, cannot be fully determined by the state of delayed luminescence, which is a relaxation from an excited state and is not stationary by definition. In other words, the state of light met in delayed luminescence is different from the state of light of autoluminescence because the former is time-dependent and the latter is not. Therefore, conclusions from the study of physical parameters of delayed luminescence cannot be directly used to prove parameters of autoluminescence. & \cite{Li-83,Popp-84} \\
\hline
Delayed luminescence in plants is usually interpreted as a consequence of the complex reactions involved in photosystem II~\citep{Goltsev-09}. Experimental delayed luminescence can then be reproduced using reasonable reaction constants~\citep{Guo-09}. Popp and Li used a different approach. They postulated that the intricate behaviour of the photosynthetic chain could be modelled by a simple one-dimensional harmonic oscillator with a time-dependent force:
$\ddot{x}(t)+ 2 \mu(t) \dot{x}(t)+\omega_0^2 x(t)=0$.
They remove the term in $\dot{x}$ by writing
$x(t)=\exp\big(-\int_{0}^t \mu(\tau)d \tau\big) y(t)$,
where $y$ satisfies the equation
$\ddot{y}+ (\omega_0^2-\mu^2 - \dot{\mu}) y=0$.
Without any reasonable justification, they further postulate that the oscillating part $y(t)$ should have a constant frequency. This gives us the equation $\mu^2 + \dot{\mu}=\omega^2$, so that $y$ oscillates
with constant frequency $\sqrt{\omega_0^2 - \omega^2}$. The basic solutions of this equation are
$\mu(t)=\omega\tanh(\omega t + \mu_0)$, so that $x(t)=e^{\pm i \omega t} \cosh \mu_0/\cosh(\mu_0+\omega t)$.
Without any justification, Popp and Li completely dismiss these general solutions and choose the very special
$\mu(t)=-\omega\tan(\omega t + \mu_0)$, which corresponds to $\omega=0$. & \cite{Li-83,Popp-84,Popp-Li-92,Popp-93} \\
\hline
Popp and Yan use the above mentioned special solution and try to get a coherent-state model of delayed luminescence. The solution of the problem does not satisfy them (it is not compatible with experiments) and the desired solution is achieved in the procedure which unfortunately involves several mathematical errors - see ~\citep{Salari-11} for the detailed critical treatment. & \cite{Popp-02,Yan-05} \\
\hline
Numerous evidences are provided for a Poisson distribution of biological UPE by showing two examples where $\langle n\rangle \simeq \langle (\Delta n)^2 \rangle$ for cucumber seedlings with and without poisoning by acetone. However, the non-poisoned case is not compatible with the value previously reported~\citep{Popp-81}. The other measurements of the previous reference~\citep{Popp-81}, that are not compatible with a Poisson distribution, are not mentioned. He admits that, for a large number of degrees of freedom $M$, a chaotic field would also have $\langle n\rangle \simeq \langle (\Delta n)^2 \rangle$. 
However, this point of view is dismissed because ``we found an extremely strong mode-coupling indicating that $M$ is of order 1.'' As a reference, ref.~\citep{Popp-81} is cited where this statement or similar supporting it cannot be found.& \cite{Popp-86}\\
\hline
\emph{Coherence hypothesis} is stated, which claims in very general terms that ``biophotons are released from a fully coherent electromagnetic field which serves as a basis for communication in living tissues''~\cite[p.~577]{Popp-88}. They show measurement of cucumber seedlings with smaller (but variable) values of $\delta$. They argue that the statistics alone does not prove
coherence, but that the temperature dependence, the transparency of biological materials and the hyperbolic decay of luminescence do~\cite[p.~581]{Popp-88}. These statements are not supported by any rigorously convincing proof. They set up a simple model to describe the emission of coherent light by DNA. & \cite{Popp-88}\\
\hline
Several strong (but largely unfounded) statements are made in this reference. For example, it is written that the  phase is completely determined in a coherent state~\cite[p.~147]{Popp-89}, whereas in fact the variance of the phase is $1/(2\sqrt{\langle n\rangle})$~\cite[p.~196]{Loudon-00},which can be very large for the low intensity of UPE. & \cite{Popp-89}\\
\hline
Further examples of statements which seem to be of conclusive nature but are unfounded: ``While spontaneous chemiluminescence cannot sensitively depend on biological and physiological processes, like the cell-cycle, growth phases, differentiation, enzymatic activity, conformational changes of DNA, the external temperature, and weak external perturbations, the opposite behaviour is expected for a coherent field, since it is modulated by any small change of the boundary conditions, including all the environmental and internal factors.''~\cite[p.~148]{Popp-89} and ``As far as results are available, there is no indication for the validity of hypothesis 1, the chaos theory, but complete support of hypothesis 2, namely the coherence theory of biophotons.''~\cite[p.~148]{Popp-89}
Equation $\langle x \ddot{x}\rangle=(1+\kappa) \langle {\dot{x}}^2\rangle$~\cite[p.~160]{Popp-89} is solved as if it was the equation $ x \ddot{x}=(1+\kappa) {\dot{x}}^2$. This is not correct in general.  & \cite{Popp-89}\\
\hline
Several novelties are introduced in a review paper~\citep{Popp-92}. Factorial moments~\cite[p.~71]{Arecchi-69} are used to describe photocount statistics, new experimental photocount distributions are presented and an optical biocommunication experiment between \emph{Gonyaulax polyedra} is described without giving closer details on experimental protocols. In the same book, Popp makes several speculative statements about ``evolution as the expansion of coherent states''~\citep{Popp-92-evolution}, Popp and Li see ``hyperbolic relaxation as a sufficient condition of a fully coherent ergodic field''~\citep{Popp-Li-92}. \cite{Li-92} postulates super-radiance in DNA and mentions squeezed states. & \cite{bpe-book1992}\\
\hline
Optical biocommunication experiment between \emph{Gonyaulax polyedra} is ascribed to super-radiance. & \cite{Popp-94} \\
\hline
Results of coincidence counting of UPE from mungbean seedlings and an elder bush leaflet are published~\citep{Popp-98}. When a photon is registered in channel 1, the photons in channel 2
are registered during the time interval $\Delta t$. A coincidence occurs when at least one photon is detected in channel 2. For a non-stationary process, the number of random coincidences $Z_j$ in the $j$-th time interval $\Delta T_j$ is $Z_j=n_{1j} (1-P_2(\Delta t,0))$, where $P_2(\Delta t,0)$ is the probability of counting no photon in the time interval $\Delta t$ and $n_{1j}$ is the number of counts in channel 1 in the
$\Delta T_j$. For a Poissonian distribution we have $P_2(\Delta t , 0)=\ee^{-a\Delta t}$, where $a$ is determined by $\langle n\rangle = a \Delta t$.Therefore, $Z_j=n_{1j} (1-\ee^{-a \Delta t})$. Experimental results for mungbean (\emph{Vigna radiata}) seedlings and elder bush (\emph{Sambucus nigra}) leaflets are given.
Photocount statistics agree with the Poissonian distribution. 
Similar experiments were done on soybeans (\cite{Chang-98}, see also ref.~\cite{Popp-00}). Poissonian distribution is interpreted there in terms of super-radiance, although super-radiance does not generally generates Poissonian photocount statistics (see section \ref{superrad}). &  \cite{Popp-98,Chang-98,Popp-00}\\
\hline
Squeezed states were used instead of super-radiance as an attempt to describe UPE photocount statistics \citep{Popp-02-non-classical}. Squeezed states they used are not as general as the ones of \cite{Bajpai-98-AMC}. They are of the form $|\alpha,r\rangle = D(\alpha) S(r) |0\rangle$. If we compare with Eq.~(21.3-1) of \cite[p.~1038]{MandelWolf} we see that $z=r \ee^{\mathrm{i}\theta}$ is real, so that $\theta=0$,  $\alpha=v$ \cite[p.~1042]{MandelWolf} and $|\alpha,r\rangle = |(v,z)\rangle$ is an ideal squeezed state (21.4-2) of \cite[p.~1042]{MandelWolf}. For Mandel and Wolf $\mu=\cosh r$ and $\nu=\ee^{\mathrm{i}\theta}\sinh r$ in Eq.~(21.3-3) of \cite[p.~1039]{MandelWolf}. The values of $p(n)$ and $p(0)$ are not correct. For example, the value of $p(n)$ given in the paper is not real when $\alpha$ is not real. But the formula is wrong even  if $\alpha$ is real. The correct form for $\alpha$ real is:
\begin{eqnarray*}
   p(n)&=& \frac{1}{n! \cosh r} \left(\frac{\tanh r}{2}\right)^n
  \ee^{-\alpha^2(1+\tanh r)}
   H_n^2\left(\frac{\alpha \ee^r}{\sqrt{\sinh 2r}}\right).
\end{eqnarray*}
Thus,
\begin{eqnarray*}
   p(0)&=& \frac{1}{\cosh r} \ee^{-\alpha^2(1+\tanh r)}.
\end{eqnarray*}
By using $\langle n\rangle = \alpha^2 + \sinh^2 r$ we obtain
\begin{eqnarray*}
 p(0)&=& \frac{1}{\cosh r} \ee^{-(\langle n\rangle -\sinh^2 r)(1+\tanh r)},
\end{eqnarray*}
which does not reduce to the expression for $p(0)$ given in \citep{Popp-02-non-classical}. Unfortunately, this incorrect formula is repeated in \citep{Chang-08,Chang-08-neuro}.  & \cite{Popp-02-non-classical}\\

\hline
Review articles with no new results and similar issues as those mentioned above. & \cite{Popp-03,Popp-03-Indian,Popp-09}\\
\hline
\end{longtable}

The research work described in this section was led by the working hypothesis that the coherence is the fundamental principle responsible for functioning of biological systems. Fine experimental setups were built, clever experiments with very interesting results performed, but there are several methodological drawbacks: (i) experiments are not described in detail, (ii) surprising experimental results (for example the concentration dependence of UPE in \emph{Daphnia}) are not repeated with many other samples or other experimental setups, (iii) data that do not agree with the coherence interpretation are dismissed, (iv) alternative interpretations are not seriously considered, (v) oversimplified models are used instead of realistic biophysical ones, (vi) mathematical errors in the articles. While the ideas presented inspired many researchers, incorrect and controversial interpretations of the data brought the subject of ``biophotons'' into disrepute.

\subsubsection{Squeezed states of ultra-weak photon emission ?}
The squeezed state of light provides a flexible shape to fit the UPE photocount statistics because this state is based on four independent parameters ($|\alpha|, \phi, r, \theta$), see Fig. \ref{fig-squeezing}. This interesting model was first introduced by R.P. Bajpai \cite{Bajpai-98-AMC}. However, the fact that this model fits experimental data does not mean that UPE is in a squeezed state. We saw that a Poisson distribution can be obtained from a coherent state but also from many other (classical or quantum) states of light. The same is true for the distribution given by squeezed state. For example, Mandel and Wolf notice that for certain values of the squeezed-state parameters, the Mandel parameter takes positive (\emph{i.e.} classical) values~\cite[p.~1051]{MandelWolf}. Even if restricted to Fock space, the number of states of light is immensely larger than the number of photocount distributions. Thus, it is generally impossible to deduce a state of light from a photocount distribution. Higher order correlation functions must be measured. As in the case of previous section, interesting experimental results are somewhat spoiled by speculative interpretations~\citep{Bajpai-00,Bajpai-09}.

%%%%%%%%%%%%%%%%%
\begin{longtable}[h]{|p{9cm} | p{3cm} |}
\caption{Chronologically ordered assessment and brief description of publications focused on squeezed states \label{table:bajpai}}\\

%\begin{table}[h]
%\begin{center}
%\begin{tabular}{|l|c|}
\hline
Statements, issues & Main references\\
\hline
Paper \cite{Bajpai-98-AMC} presents an analysis of UPE photocount distribution with squeezed states $|\beta,\mu_t,\nu_t\rangle$  (or more precisely, Yuen's two-photon coherent states~\cite[p.~1046]{MandelWolf}) instead of standard coherent states. The relation with Mandel and Wolf is $\mu=\mu$, $\nu=\nu$, $w=\beta$. Authors impose a hyperbolic decay $\lambda(t)=\lambda_0/(1+\lambda_0 t)$ and they find a time-dependent pseudo-annihilation operator $b(t)=\mu_t \hat{a} + \nu_t \hat{a}^\dagger$, with $\mu_t$ and $\nu_t$ explicitly given. From this, they compute $n(t)$ and remove the oscillatory terms. They get an expression $B_0+ B_1/(1+\lambda_0 t) + B_2/(1+\lambda_0 t)^2$ but they find that $B_1=0$. They also calculate the Mandel factor $Q = \langle \Delta (\Delta n)^2 \rangle - \langle n \rangle$. They find it non-zero but small.
%Of course ``$B_0$ and $B_2$ are holistic in character''.
They consider earlier experiments on flowers of \emph{Tagetes patula}. They had fitted the fluorescence decay with a sum of two exponentials, but they say that using their new formula gives also a good fit. & \cite{Bajpai-98-AMC}\\
\hline

In \cite{Bajpai-98}, coincidence measurements made with Popp's experimental setup~\citep{Popp-94} using two photomultipliers are presented. Experiments were carried out on ``leaves of different sizes from different plants''.  Some experimental results are given but the author does not specify for which sample. The paper ends with the idea that, from the evolutionary point of view, ``the advantages of using squeezed light were too overwhelming". & \cite{Bajpai-98}\\
\hline

Bajpai \cite{Bajpai-99} argues that the ``inadequacy of the conventional framework to describe a biophoton signal is easy to demonstrate''. Then, the ``separate identity of sub-units and the independence of de-excitations give rise to the thermal nature of photons and exponential decay character of the signal''. This is generally not correct. The statement in the conclusion ``The signal was, therefore, coherent for 5 hr" does not seem to be substantiated. & \cite{Bajpai-99}\\
\hline

Here \cite{Bajpai-03}, Bajpai accepts another definition of coherent states and the real $r$ is replaced by the complex number $\xi$, which is the same as $z=r \ee^{\mathrm{i}\theta}$ in Eq.~(21.3-1) of \cite{MandelWolf}[p.~1038]. From this paper on, this definition of squeezed coherent states is kept. However, his value of $p_0(\langle n\rangle)$ is wrong in this paper, since it would be complex when $\alpha$ is complex. The denominator is also wrong.

The correct result is 
  \begin{eqnarray*}
  p_0(\langle n\rangle) &=& \ee^{-|\alpha|^2} 
  \frac{\ee^{- \Re(\alpha^2 \ee^{-\mathrm{i}\theta}) \tanh r}}
    {\cosh r}
\\&=&
  \ee^{-\langle n \rangle} 
  \frac{\ee^{\sinh^2 r - \Re(\alpha^2 \ee^{-\mathrm{i}\theta}) \tanh r}}
    {\cosh r},
  \end{eqnarray*}
where we used
$\langle n \rangle = |v|^2+|\nu|^2=|\alpha|^2 + \sinh^2 r$ from Eq.~(21.4-10) of \cite[p.~1044]{MandelWolf}.
  The value of $p(n)$ is obtained from Eq.~(21.5-25) of \cite{MandelWolf}[p.~1050] and the value of $\langle n|\alpha \xi\rangle$ is obtained from   $\langle n| {[}\mu,\nu;w{]}\rangle$ of Eq.~(21.5-24) of \cite{MandelWolf}[p.~1050] by the substitution $\mu=\cosh r$, $\nu=\ee^{\mathrm{i}\theta} \sinh r$ and $w=\alpha \cosh r + \alpha^* \ee^{\mathrm{i}\theta}\sinh r$. This is explained in Eq.~(21.4-5) of \cite{MandelWolf}[p.~1042]. & \cite{Bajpai-03}\\
   
\hline 
Lichen \emph{Parmelia tinctorum} is measured in \cite{Bajpai-04,Bajpai-05} since ``Lichen, because of its very slow growth or decay, is a suitable system for making repeated experiments''. Results are analyzed using squeezed state distributions. In these and the following works by Bajpai, the formulas for the photocount distribution of squeezed states are correct. It is stated that ``Since a photon signal of quantum nature emanates from a quantum state the biophoton emitting parts of a living system must remain in a pure quantum state''. This is not true in general. The fact is that a classical source generates a coherent state of the photon field, although the source is not quantum at all. Even squeezed states can be generated by classical currents in the non-linear regime. The only requirement is the presence of a quadratic term (in the photon creation and annihilation operators) in the interaction Hamiltonian~\citep{Zhang-00}. & \cite{Bajpai-04,Bajpai-05}\\
\hline

Photocount statistics of \emph{Parmelinella wallichiana} is measured and fitted by a squeezed state statistics. & \cite{Bajpai-05-PLA}\\
\hline
Time-dependent squeezed states with a time dependence giving a density $n(t)=\sum_{i=0}^2 B_i (t_0+t)^{-i}$ are considered. Then, for $t\to\infty$, $p(n)$ is a distribution coming from a squeezed state. 
 \begin{eqnarray*}
  \hat{H} &=& f(t)\hat{p}^2 +  g(t) \omega^2 \hat{q}^2.
   \end{eqnarray*}
 We write $q=\beta \hat{a} + \beta^\dagger \hat{a}^\dagger$ and  $p=\alpha \hat{a} + \alpha^\dagger \hat{a}^\dagger$, where $\hat{a}^\dagger$   and $\hat{a}$ are creation and annihilation operators. The Hamiltonian becomes
  \begin{eqnarray*}
   \hat{H} &=& 2f_1 \hat{a}^\dagger \hat{a} + f_2^* (\hat{a}^\dagger)^2 + f_2 \hat{a}^2
  + f_1, 
  \end{eqnarray*}
where 
  \begin{eqnarray*}
   f_1 &=& |\alpha|^2 f + |\beta|^2 \omega^2 g,\\
   f_2 &=& \alpha^2 f + \beta^2 \omega^2 g.
  \end{eqnarray*}
  This type of Hamiltonian was investigated in detail also elsewhere~\citep{Yuen-76,Liu-92}.
  Delayed luminescence from \emph{Parmelinella wallichiana} is measured. One of the fits is displayed, the noise is very large. 

& \cite{Bajpai-07}\\
\hline
An interesting and critical survey of previous measurements. Additionally, new measurements of \emph{Xanthoria parietina} and a rather good fit of the data using squeezed-state distributions are presented. Speculative statements: ``A holistic property is correctly described only in the quantum framework'',  ``The photon signal remained in its squeezed state at least for 5 hr'', ``emission of photon signal in squeezed state is a characteristic property of living systems''. & \cite{Bajpai-08-Xanthoria}\\

\hline
In \cite{racine2013}, authors introduced second order correlation function at zero time lag $g^{(2)}(\tau = 0)$ to estimate non-classicality of UPE. By varying the detector response time (or bin size) around the period of expected coherence time one can estimate the coherence time and $g^{(2)}(0)$ \citep{assmann2012}. $g^{(2)}(0) < 1$ is a signature of non-classical (quantum nature) of the light \cite[Fig.~5.21,p.~229, p.~249-250]{Loudon-00}.
Racine and Bajpai \cite{racine2013} estimated $g^{(2)}(0)$ from the Fano factor of measured UPE signal and compensated for the detector background. In Fig.~2 in \cite{racine2013}, $g^{(2)}(0) = 1$ and shows no variation across different bin sizes for hydrogen peroxide induced UPE signal from human hand. Authors claim there that there are hints to quantum behavior of UPE ($g^{(2)}(0) < 1$) but the data provided do not fully justify that. Fluctuation of Fano factor sometimes fall under the value of 1 (again signature of quantum states of light) for certain bin sizes, but considering the shape of the Fano factor curve as whole \cite[Fig.~2]{racine2013} this could be also attributed to fluctuation and error of the measurement. Nevertheless, these interesting data should be reproduced and thoroughly verified.& \cite{racine2013}\\
\hline

\hline
\end{longtable}

\section{Conclusion and perspectives}
We reviewed practically all available literature on the statistical properties of UPE. There are several high quality works on the level of standard quantum optics literature and provide provide analysis of UPE in terms of chaotic light field. In contrast, there are numerous papers which contain claims about coherent and squeezed states of UPE. However, only incorrect argumentation and data interpretation or indirect anecdotal evidence is largely presented to support these claims. 

The conclusion of our review is that while the phenomenon of UPE from biological systems can be considered experimentally well established, no reliable evidence for the coherence or nonclassicality of UPE was actually achieved up to now. The presence of coherence seems to follow from a straightforward reasoning: a living organism must be in some coherent state because it is obviously not in thermal equilibrium~\citep{Giudice-05}. However, the actual situation is subtle. On the one hand a thermal source can emit partially coherent light, even close to the source~\citep{Greffet-02}, and independent thermal sources can produce two-photon interference~\citep{Zhai-06}\footnote{Note that two-photon interference is not the interference of two photons~\citep{Pittman-96}}.
On the other hand the organization required to maintain life has no \emph{a priori} reason to imply that UPE is in a coherent state. Moreover, thermal states and coherent states are two extremes of a very broad range of possible states of light. 
What we would need is to actually \emph{measure} the coherence length and time of UPE. The extremely long UPE coherence times (10~days\footnote{``A reasonable coherence time is the lifetime of cell organelles (for instance, mitotic figures) of about ten days''~\cite[p.~59]{Popp-09}.}, 5~hr\footnote{``The signal was, therefore, coherent for 5 hr"~\citep{Bajpai-99}.}) proposed by some authors seem to be completely off the mark. It is remarkable that, except for a few exceptions~\citep{Letokhov-03,Salari-11}, the physical community did not provided almost any critique of these extraordinary claims.

Although the role of coherent processes in biology, in particular quantum coherence, cannot be dismissed in general \citep{parson2007,wolynes2009,Engel-07,Shelly-08}, it needs to be emphasized that the research work published until now does not provide any generally accepted proof for coherence of biological ultra weak photon emission according to the physical definitions (see section \ref{coherence}).

Perspectively, standard methods in quantum optics can deliver more reliable information on coherence and statistical properties of UPE of living systems. Coherence parameters (coherence time, coherence length) could be quantified by measuring light interference or light correlation functions \cite{MandelWolf,wolf2007}. A non-classical, \emph{i.e.} quantum nature could be assessed by using a Hanbury Brown-Twiss interferometer and measuring higher order correlation functions. However, the extremely low intensity of UPE and inherent nonstationarity of the biological signal make these experiments highly challenging.

We believe that the development of new types of photon detectors which will have properties closer to that of the ideal detectors \cite[sec.~5]{cifra2014-jppb} may bring at least partial answers to the open questions about UPE statistical properties. Such new developments include light (200 nm $-$ 3000 nm) sensor based on the cryogenically cooled microwave kinetic inductance detectors \cite{mazin2012, mazin2013}. Further futuristic possibilities of a light detection could include nondestructive detection of the presence of photons, i.e. without absorbing them, by detecting the change of the phase they incur on pre-prepared quantum state of the atom in cavity, as was recently experimentally demonstrated \cite{reiserer2013}. A promising technological direction to explore is to couple the UPE into optical fibers. Once in a fiber, the light can be easily manipulated, spectrally and spatially filtered and small low noise avalanche photodiode (APD) detectors can be used. This manipulation allows a control of the number of the modes which can enter the detector. It has been demonstrated already 20 years ago, that in spite its low intensity UPE can be coupled to an optical fiber and detected by a liquid nitrogen cooled Si-APD \cite{isoshima1995}. Actually, fiber optics and APD detectors based setups are a standard in quantum optics experiments.

Apart from the quantum statistical properties, there are indications that other signal properties of biological UPE stemming from dynamics underlying chemical reactions \citep{iranifam2010, voeikov4, voeikov5, vanWijk2011, scholkmann2011} may be also of interest. Biological processes are naturally oscillatory, complex (chaotic) and fractal. Thus, suitable methods adapted from statistical physics and very carefully used for other biological signals to uncover ``hidden information''\citep{goldberger2002} may be also used to analyze the UPE signals.

\section*{Acknowledgement}
Ch.B. thanks to Reinhard Honegger for his kind explanations of the relation between coherent states and factorizing correlations. Despite being critical to their work, M.C. acknowledges F.-A. Popp for introducing him to the field of biological ultra-weak photon emission and R. P. Bajpai for extensive discussions and inspirations. M.C. and M.N. are financially supported by the Czech Science Foundation, grant no. GP13-29294S.

\section*{Authors' contribution}
M.C. and Ch.B. conceived the research. M.N. and M.C. performed the calculations and analyzed the data. M.C., Ch.B. and OK wrote the paper. All authors read and approved the final manuscript.

\section*{Appendix A: Other works on statistical properties of UPE with speculative interpretations}
\label{append-specul}
Several more authors indulged themselves in speculations about nature of the light emitted from biosystems. We list their relevant publication here for the sake of completeness. Gu has strong theoretical background in quantum optics. In \cite{Gu-92}, he describes a three-level system as the emitter of UPE. Super-radiance and a model involving the sum of two coherent  states is He introduced in~\citep{Gu-95}. In \cite{Gu-98}, Gu discusses non-classical light and asks the question ``are there nonclassical effects in biological systems?'' (p.~301). He recalls the non-classical aspect of sub-Poissonian photon distribution from \emph{Gonyaulax polyedra}~\citep{Popp-92} and higher order coherence in mungbean seedlings~\cite[p.~1272]{Popp-94}. Further, he states that biophotons may be emitted by standing vibrational waves in DNA. Gu considers the interaction of a single mode of the biophoton field with a phonon reservoir. He considers a Schr\"odinger cat initial state (healthy or ill, yin or yang). These theoretical considerations are apparently not used in the paper. Photon statistics of a piece of leaf of banyan tree (probably \emph{Ficus elastica}) is reported. The photon distribution, its variance and entropy are given for the leaf and a radiator. Gu further compares the variance and entropy observed during the delayed luminescence and autoluminescence phases, and observes that they are similar. The normalized variance for the leaf and the radiator are 1.26 and 2.20, respectively. He compares the value for the leaf (1.26) to the value $g^{(2)}(0)=1.2$ obtained in \cite[p.~83]{Gu-92}. Other measurements give values much closer to 1. An important point is that author does not measure $\langle n^2 \rangle$, he calculates it from the distribution $p(n)$. Thus, he cannot observe non-classical effects because $p(n)$ is always positive. Finally, Gu compares the variance and entropy for traditional and genetically modified soybeans.
Extensive theoretical work of Gu is covered in his book \cite{gu2003}.

Kun \cite{Kun-09} considers the single-mode coherent states corresponding to parameters $\alpha$ and $-\alpha$. They make the same mistake as Popp \cite{Popp-02} and consider that $\omega(t)$ and $\beta(t)$ can be chosen independently. This is wrong because the time dependence of $n(t)$ is determined by $\omega$ and $f$~\citep{Salari-11}.

Chang \cite{Chang-08} describes coincidence counting experiments. Distributions were measured for Dinoflagellates, chicken embryos, fireflies \emph{Lampyridae}. She discusses Popp's hypothesis that biophotons come from DNA: ``DNA excimer radiation is based on the same principle as laser radiation''. She pushes this hypothesis very far: ``During gene transcription the long distance regulative functions may be performed by biophotons.'' ``Presumably one of the neurofilament's functions is to act as transmission channels for photon signals.'' ``The biophoton fields are in coherent and squeezed states suggesting that over a long period of life evolution livings learned how to use quantum mechanism to regulate themselves.'' Another published paper~\citep{Chang-08-neuro} is of similar nature as the previous one. Reference \cite{Lozneanu-08} is another speculative paper, see for instance one statement: ``The emission of biophotons becomes coherent when the minuscule electric double layers start their moving state at the same moment.''

\newpage
\section*{Appendix B: Table of UPE photocount statistics experiments}

\begin{longtable}[h]{|l|l|p{3.6cm}|}
%\begin{tabular}{|l|l|p{3.6cm}|}
\hline
{\bf Sample from} & & {\bf References}\\
\hline
\hline

{\bf Chemicals} & & \\
\hline 
luminol & C$_8$H$_7$N$_3$O$_2$ & \cite{Shen-93}\\
\hline
polystyrene & (C$_8$H$_8$)$_n$   & \cite{Popp-92}\\
\hline
9,10-diphenylanthracene & C$_{26}$H$_{18}$ & \cite{Collinson-95}\\
\hline

%\multicolumn{3}{|c|}{Chemicals} 
{\bf Prokaryotes} & & \\
\hline
symbiotic bacteria & \emph{Photobacterium phosphoreum} & \cite{Kobayashi-98}\\
\hline

nitrogen-fixating symbiont & \emph{Bradyrhizobium japonicum} & \cite{Shen-93}\\
\hline

{\bf Eukaryotes, unicellular} & & \\
\hline
``umbrella'' or cap algae & \emph{Acetabularia acetabulum}  & \cite{Popp-92}\\
\hline

dinoflagellate & \emph{Prorocentrum elegans} & \cite{Popp-92}\\
\hline

dinoflagellate & \emph{Gonyaulax polyedra} & \cite{Popp-92,Gu-95,%
   Chang-08}\\
\hline

slime mold (also multicellular) & \emph{Dictyostelium discoideum} & \cite{Kobayashi-00}\\
\hline

{\bf Algea-mushroom symbiont} & & \\
\hline
lichen & \emph{Parmelia physodes} & \cite{SchirmacherPhD} \\
\hline
lichen & \emph{Parmelia tinctorum} & \cite{Bajpai-04,Bajpai-05} \\
\hline
lichen & \emph{Parmelinella wallichiana} & \cite{Bajpai-05-PLA,Bajpai-07} \\
\hline

lichen on a tree bark & \emph{Xanthoria parietina} & \cite{Bajpai-08-Xanthoria}\\
\hline

{\bf Plants} & & \\
\hline

silver fir twig & \emph{Abies alba} & \cite{SchirmacherPhD} \\
\hline

arabica coffee grains & \emph{Coffea arabica}  & \cite{Gallep-04}\\
\hline
robusta coffee grains & \emph{Coffea canephora}  & \cite{Gallep-04}\\
\hline

cucumber seedlings & \emph{Cucumis sativus} & \cite{Popp-81,Shen-93}\\
\hline

cucumber & \emph{Cucumis sativus} & \cite{Gu-95}\\
\hline

elder bush leaflet & \emph{Sambucus sp.} & \cite{Popp-98}\\
\hline

banyan tree leaf & \emph{Ficus microcarpa} & \cite{SchirmacherPhD} \\
\hline

gum tree leaf & \emph{Ficus elastica} & \cite{Gu-98}\\
\hline

mungbean seedlings & \emph{Phaseolus aureus} & \cite{Shen-93,Popp-94,Popp-98}\\
\hline

purple plum leaf & \emph{Prunus cerasifera} `Nigra' & \cite{SchirmacherPhD} \\
\hline

oak leaf & \emph{Quercus robur} & \cite{SchirmacherPhD} \\
\hline

soybean seedlings & \emph{Glycine max} & \cite{Popp-92,Popp-94}\\
\hline

soybeans & \emph{Glycine max} &  \cite{Chang-98}\\
\hline

stinging nettle leaf & \emph{Urtica dioica} & \cite{SchirmacherPhD} \\
\hline

{\bf Animal} & & \\
\hline
waterfleas (Crustacean) & \emph{Daphnia sp.} & \cite{Popp-92-evolution,Gu-95,Gallep-07}\\
\hline

fireflies (Insects) & \emph{Lampyridae} & \cite{Chang-08}\\
\hline

thailand firefly (Insects) & \emph{Lampyridae} & \cite{Popp-92}\\
\hline

chicken embryo, brain & \emph{Gallus gallus domesticus} & \cite{Chang-08}\\
\hline

{\bf Man} & & \\
\hline
body & \emph{Homo sapiens sapiens} & \cite{Wijk-06,bajpai2013}\\
\hline
body of meditating subjects & \emph{Homo sapiens sapiens} & \cite{Wijk-08}\\
\hline
hand of a multiple sclerosis patient
   & \emph{Homo sapiens sapiens} & \cite{Bajpai-08}\\
\hline
hands & \emph{Homo sapiens sapiens} & \cite{Wijk-10,racine2013}\\
\hline

%\end{tabular}
%\end{table}
\end{longtable}

%\section*{References}

\bibliographystyle{model1a-num-names}
\bibliography{merged_mybib}

\begin{thebibliography}{181}
\expandafter\ifx\csname natexlab\endcsname\relax\def\natexlab#1{#1}\fi
\providecommand{\url}[1]{\texttt{#1}}
\providecommand{\href}[2]{#2}
\providecommand{\path}[1]{#1}
\providecommand{\DOIprefix}{doi:}
\providecommand{\ArXivprefix}{arXiv:}
\providecommand{\URLprefix}{URL: }
\providecommand{\Pubmedprefix}{pmid:}
\providecommand{\doi}[1]{\href{http://dx.doi.org/#1}{\path{#1}}}
\providecommand{\Pubmed}[1]{\href{pmid:#1}{\path{#1}}}
\providecommand{\bibinfo}[2]{#2}
\ifx\xfnm\relax \def\xfnm[#1]{\unskip,\space#1}\fi
%Type = Article
\bibitem[{Cifra and Posp{\'\i}{\v{s}}il(2014)}]{cifra2014-jppb}
\bibinfo{author}{M.~Cifra}, \bibinfo{author}{P.~Posp{\'\i}{\v{s}}il},
  \bibinfo{journal}{Journal of Photochemistry and Photobiology B: Biology}
  \bibinfo{volume}{139} (\bibinfo{year}{2014}) \bibinfo{pages}{2--10}.
%Type = Article
\bibitem[{Posp{\'\i}{\v{s}}il et~al.(2014)Posp{\'\i}{\v{s}}il, Prasad, and
  R{\'a}c}]{pospisil2014}
\bibinfo{author}{P.~Posp{\'\i}{\v{s}}il}, \bibinfo{author}{A.~Prasad},
  \bibinfo{author}{M.~R{\'a}c}, \bibinfo{journal}{Journal of Photochemistry and
  Photobiology B: Biology} \bibinfo{volume}{139} (\bibinfo{year}{2014})
  \bibinfo{pages}{11--23}.
%Type = Article
\bibitem[{Havaux et~al.(2006)Havaux, Triantaphylid{\`e}s, and
  Genty}]{havaux2006}
\bibinfo{author}{M.~Havaux}, \bibinfo{author}{C.~Triantaphylid{\`e}s},
  \bibinfo{author}{B.~Genty}, \bibinfo{journal}{Trends in Plant Science}
  \bibinfo{volume}{11} (\bibinfo{year}{2006}) \bibinfo{pages}{480--484}.
%Type = Article
\bibitem[{Quickenden and {Que Hee}(1974)}]{quickenden1974}
\bibinfo{author}{T.~Quickenden}, \bibinfo{author}{S.~S. {Que Hee}},
  \bibinfo{journal}{Biochemical and Biophysical Research Communications}
  \bibinfo{volume}{60} (\bibinfo{year}{1974}) \bibinfo{pages}{764--770}.
%Type = Article
\bibitem[{Cadenas et~al.(1980)Cadenas, Boveris, and Chance}]{cadenas2}
\bibinfo{author}{E.~Cadenas}, \bibinfo{author}{A.~Boveris},
  \bibinfo{author}{B.~Chance}, \bibinfo{journal}{Biochemical Journal}
  \bibinfo{volume}{186} (\bibinfo{year}{1980}) \bibinfo{pages}{659--667}.
%Type = Article
\bibitem[{Cohen and Popp(1997)}]{cohen1997}
\bibinfo{author}{S.~Cohen}, \bibinfo{author}{F.-A. Popp},
  \bibinfo{journal}{Journal of Photochemistry and Photobiology B: Biology}
  \bibinfo{volume}{40} (\bibinfo{year}{1997}) \bibinfo{pages}{187--189}.
%Type = Article
\bibitem[{Devaraj et~al.(1997)Devaraj, Usa, and Inaba}]{devaraj1997}
\bibinfo{author}{B.~Devaraj}, \bibinfo{author}{M.~Usa},
  \bibinfo{author}{H.~Inaba}, \bibinfo{journal}{Current Opinion in Solid State
  and Materials Science} \bibinfo{volume}{2} (\bibinfo{year}{1997})
  \bibinfo{pages}{188--193}.
%Type = Inproceedings
\bibitem[{Popp(1986)}]{Popp-86}
\bibinfo{author}{F.-A. Popp}, in: \bibinfo{editor}{C.~W. Kilmister} (Ed.),
  \bibinfo{booktitle}{Disequilibrium and Self-Organization},
  \bibinfo{publisher}{Reidel Publishing Company}, \bibinfo{address}{Dordrecht},
  \bibinfo{year}{1986}, pp. \bibinfo{pages}{207--30}.
%Type = Article
\bibitem[{Popp et~al.(1994)Popp, Gu, and Li}]{Popp-94}
\bibinfo{author}{F.-A. Popp}, \bibinfo{author}{Q.~Gu}, \bibinfo{author}{K.-H.
  Li}, \bibinfo{journal}{Mod. Phys. Lett. B} \bibinfo{volume}{8}
  (\bibinfo{year}{1994}) \bibinfo{pages}{1269--96}.
%Type = Article
\bibitem[{Popp(2009)}]{Popp-09}
\bibinfo{author}{F.-A. Popp}, \bibinfo{journal}{Electromagn. Biol. Medic.}
  \bibinfo{volume}{28} (\bibinfo{year}{2009}) \bibinfo{pages}{53--60}.
%Type = Article
\bibitem[{Popp(2003)}]{Popp-03-Indian}
\bibinfo{author}{F.-A. Popp}, \bibinfo{journal}{Indian J. Exp. Biol.}
  \bibinfo{volume}{41} (\bibinfo{year}{2003}) \bibinfo{pages}{391--402}.
%Type = Article
\bibitem[{Popp et~al.(2002)Popp, Chang, Herzog, Yan, and
  Yan}]{Popp-02-non-classical}
\bibinfo{author}{F.-A. Popp}, \bibinfo{author}{J.~J. Chang},
  \bibinfo{author}{A.~Herzog}, \bibinfo{author}{Z.~Yan},
  \bibinfo{author}{Y.~Yan}, \bibinfo{journal}{Phys. Lett. A}
  \bibinfo{volume}{293} (\bibinfo{year}{2002}) \bibinfo{pages}{98--102}.
%Type = Inproceedings
\bibitem[{Bajpai(1998)}]{Bajpai-98}
\bibinfo{author}{R.~P. Bajpai}, in: \bibinfo{editor}{J.-J. Chang},
  \bibinfo{editor}{J.~Fisch}, \bibinfo{editor}{F.-A. Popp} (Eds.),
  \bibinfo{booktitle}{Biophotons}, \bibinfo{publisher}{Kluwer Academic
  Publishers}, \bibinfo{address}{Dordrecht}, \bibinfo{year}{1998}, pp.
  \bibinfo{pages}{323--39}.
%Type = Inproceedings
\bibitem[{Bajpai(2007)}]{Bajpai-00}
\bibinfo{author}{R.~P. Bajpai}, in: \bibinfo{editor}{L.~V. Beloussov},
  \bibinfo{editor}{F.-A. Popp}, \bibinfo{editor}{V.~L. Voeikov},
  \bibinfo{editor}{R.~van Wijk} (Eds.), \bibinfo{booktitle}{Biophotonics and
  Coherent Systems}, \bibinfo{publisher}{Moscow University Press},
  \bibinfo{address}{Moscow}, \bibinfo{year}{2007}, pp.
  \bibinfo{pages}{135--40}.
%Type = Article
\bibitem[{Ku\v{c}era and Cifra(2013)}]{kucera2013}
\bibinfo{author}{O.~Ku\v{c}era}, \bibinfo{author}{M.~Cifra},
  \bibinfo{journal}{Cell Communication and Signaling} \bibinfo{volume}{11}
  (\bibinfo{year}{2013}) \bibinfo{pages}{87.1--87.8}.
%Type = Book
\bibitem[{Pokorn\'{y} and Wu(1998)}]{pokorny1}
\bibinfo{author}{J.~Pokorn\'{y}}, \bibinfo{author}{T.-M. Wu},
  \bibinfo{title}{Biophysical Aspects of Coherence and Biological Order},
  \bibinfo{publisher}{Academia, Praha, Czech Republic; Springer, Berlin -
  Heidelberg - New York}, \bibinfo{year}{1998}.
%Type = Article
\bibitem[{Prasad et~al.(2014)Prasad, Rossi, Lamponi, Posp{\'\i}{\v{s}}il, and
  Foletti}]{prasad2014}
\bibinfo{author}{A.~Prasad}, \bibinfo{author}{C.~Rossi},
  \bibinfo{author}{S.~Lamponi}, \bibinfo{author}{P.~Posp{\'\i}{\v{s}}il},
  \bibinfo{author}{A.~Foletti}, \bibinfo{journal}{Journal of Photochemistry and
  Photobiology B: Biology} \bibinfo{volume}{139} (\bibinfo{year}{2014})
  \bibinfo{pages}{47--53}.
%Type = Article
\bibitem[{Scholkmann et~al.(2013)Scholkmann, Fels, and Cifra}]{scholkmann2013}
\bibinfo{author}{F.~Scholkmann}, \bibinfo{author}{D.~Fels},
  \bibinfo{author}{M.~Cifra}, \bibinfo{journal}{American journal of
  translational research} \bibinfo{volume}{5} (\bibinfo{year}{2013})
  \bibinfo{pages}{586}.
%Type = Article
\bibitem[{Cifra et~al.(2011)Cifra, Farhadi, and Fields}]{cifra4}
\bibinfo{author}{M.~Cifra}, \bibinfo{author}{A.~Farhadi},
  \bibinfo{author}{J.~Z. Fields}, \bibinfo{journal}{Progress in Biophysics \&
  Molecular Biology} \bibinfo{volume}{105} (\bibinfo{year}{2011})
  \bibinfo{pages}{223--246}.
%Type = Article
\bibitem[{Trushin(2004)}]{trushin2004}
\bibinfo{author}{M.~Trushin}, \bibinfo{journal}{Microbiological Research}
  \bibinfo{volume}{159} (\bibinfo{year}{2004}) \bibinfo{pages}{1--10}.
%Type = Article
\bibitem[{Nikolaev(2000)}]{nikolaev2000}
\bibinfo{author}{Y.~A. Nikolaev}, \bibinfo{journal}{Microbiology (Translated
  from Mikrobiologiya)} \bibinfo{volume}{69} (\bibinfo{year}{2000})
  \bibinfo{pages}{597--605}.
%Type = Book
\bibitem[{Loudon(2000)}]{Loudon-00}
\bibinfo{author}{R.~Loudon}, \bibinfo{title}{The Quantum Theory of Light},
  \bibinfo{edition}{third} ed., \bibinfo{publisher}{Clarendon Press},
  \bibinfo{address}{Oxford}, \bibinfo{year}{2000}.
%Type = Article
\bibitem[{Kelly and Kleiner(1964)}]{Kelley-64}
\bibinfo{author}{P.~L. Kelly}, \bibinfo{author}{W.~H. Kleiner},
  \bibinfo{journal}{Phys. Rev.} \bibinfo{volume}{136} (\bibinfo{year}{1964})
  \bibinfo{pages}{A316--34}.
%Type = Book
\bibitem[{Mandel and Wolf(1995)}]{MandelWolf}
\bibinfo{author}{L.~Mandel}, \bibinfo{author}{E.~Wolf}, \bibinfo{title}{Optical
  Coherence and Quantum Optics}, \bibinfo{publisher}{Cambridge University
  Press}, \bibinfo{address}{Cambridge}, \bibinfo{year}{1995}.
%Type = Article
\bibitem[{Picinbono and Bendjaballah(2005)}]{Picinbono-05}
\bibinfo{author}{B.~Picinbono}, \bibinfo{author}{C.~Bendjaballah},
  \bibinfo{journal}{Phys. Rev. A} \bibinfo{volume}{71} (\bibinfo{year}{2005})
  \bibinfo{pages}{013812}.
%Type = Inproceedings
\bibitem[{Popp and Shen(1998)}]{Popp-98}
\bibinfo{author}{F.-A. Popp}, \bibinfo{author}{X.~Shen}, in:
  \bibinfo{editor}{J.-J. Chang}, \bibinfo{editor}{J.~Fisch},
  \bibinfo{editor}{F.-A. Popp} (Eds.), \bibinfo{booktitle}{Biophotons},
  \bibinfo{publisher}{Kluwer Academic Publishers},
  \bibinfo{address}{Dordrecht}, \bibinfo{year}{1998}, pp.
  \bibinfo{pages}{87--92}.
%Type = Book
\bibitem[{Popp and Beloussov(2003)}]{bpe-book2003}
\bibinfo{editor}{F.-A. Popp}, \bibinfo{editor}{L.~V. Beloussov} (Eds.),
  \bibinfo{title}{Integrative Biophysics – Biophotonics},
  \bibinfo{publisher}{Kluwer Academic Publishers}, \bibinfo{year}{2003}.
%Type = Inproceedings
\bibitem[{Arecchi(1969)}]{Arecchi-69}
\bibinfo{author}{F.~T. Arecchi}, in: \bibinfo{editor}{R.~J. Glauber} (Ed.),
  \bibinfo{booktitle}{Quantum Optics. International School of Physics Enrico
  Fermi}, \bibinfo{publisher}{Academic Press}, \bibinfo{address}{New York},
  \bibinfo{year}{1969}, pp. \bibinfo{pages}{57--110}.
%Type = Article
\bibitem[{Ou and Kimble(1995)}]{Ou-95}
\bibinfo{author}{Z.~Y. Ou}, \bibinfo{author}{H.~J. Kimble},
  \bibinfo{journal}{Phys. Rev. A} \bibinfo{volume}{52} (\bibinfo{year}{1995})
  \bibinfo{pages}{3126--46}.
%Type = Article
\bibitem[{Vyas and Singh(1988)}]{Vyas-88}
\bibinfo{author}{R.~Vyas}, \bibinfo{author}{S.~Singh}, \bibinfo{journal}{Phys.
  Rev. A} \bibinfo{volume}{38} (\bibinfo{year}{1988})
  \bibinfo{pages}{2423--30}.
%Type = Book
\bibitem[{Grimaldi(1665)}]{Grimaldi}
\bibinfo{author}{F.~M. Grimaldi}, \bibinfo{title}{Physico-mathesis de lumine,
  coloribus, et iride, aliisque annexis libri duo},
  \bibinfo{publisher}{Vittorio Bonati}, \bibinfo{address}{Bologna},
  \bibinfo{year}{1665}.
%Type = Book
\bibitem[{Kipnis(1991)}]{Kipnis}
\bibinfo{author}{N.~Kipnis}, \bibinfo{title}{History of the Principle of
  Interference of Light}, \bibinfo{publisher}{Birkh\"auser},
  \bibinfo{address}{Basel}, \bibinfo{year}{1991}.
%Type = Incollection
\bibitem[{Brosseau(2009)}]{Brosseau-09}
\bibinfo{author}{C.~Brosseau}, in: \bibinfo{editor}{E.~Wolf} (Ed.),
  \bibinfo{booktitle}{Progress in Optics}, volume~\bibinfo{volume}{54},
  \bibinfo{publisher}{North-Holland}, \bibinfo{address}{Amsterdam},
  \bibinfo{year}{2009}, pp. \bibinfo{pages}{149--208}.
%Type = Article
\bibitem[{Visser and Schoonover(2008)}]{Visser-08}
\bibinfo{author}{T.~D. Visser}, \bibinfo{author}{R.~W. Schoonover},
  \bibinfo{journal}{Opt. Commun.} \bibinfo{volume}{281} (\bibinfo{year}{2008})
  \bibinfo{pages}{1 -- 6}.
%Type = Article
\bibitem[{Yadav et~al.(2006)Yadav, Rizvi, and Kandpal}]{Yadav-06}
\bibinfo{author}{B.~K. Yadav}, \bibinfo{author}{S.~A.~M. Rizvi},
  \bibinfo{author}{H.~C. Kandpal}, \bibinfo{journal}{J. Opt. A: Pure Appl.
  Opt.} \bibinfo{volume}{8} (\bibinfo{year}{2006}) \bibinfo{pages}{72--6}.
%Type = Article
\bibitem[{Agarwal et~al.(2005)Agarwal, Dogariu, Visser, and Wolf}]{Agarwal-05}
\bibinfo{author}{G.~S. Agarwal}, \bibinfo{author}{A.~Dogariu},
  \bibinfo{author}{T.~D. Visser}, \bibinfo{author}{E.~Wolf},
  \bibinfo{journal}{Opt. Lett.} \bibinfo{volume}{30} (\bibinfo{year}{2005})
  \bibinfo{pages}{120--122}.
%Type = Article
\bibitem[{Mujat et~al.(2004)Mujat, Dogariu, and Wolf}]{Mujat-04}
\bibinfo{author}{M.~Mujat}, \bibinfo{author}{A.~Dogariu},
  \bibinfo{author}{E.~Wolf}, \bibinfo{journal}{J. Opt. Soc. Am. A}
  \bibinfo{volume}{21} (\bibinfo{year}{2004}) \bibinfo{pages}{2414--7}.
%Type = Article
\bibitem[{Tervo and Turunen(2009)}]{Tervo-09}
\bibinfo{author}{J.~Tervo}, \bibinfo{author}{J.~Turunen},
  \bibinfo{journal}{Opt. Lett.} \bibinfo{volume}{34} (\bibinfo{year}{2009})
  \bibinfo{pages}{1001}.
%Type = Article
\bibitem[{Jones(1979)}]{Jones-79}
\bibinfo{author}{A.~G. Jones}, \bibinfo{journal}{J. Geophys.}
  \bibinfo{volume}{45} (\bibinfo{year}{1979}) \bibinfo{pages}{223--9}.
%Type = Book
\bibitem[{Reiter and Gabor(1928)}]{Gabor-28}
\bibinfo{author}{T.~Reiter}, \bibinfo{author}{D.~Gabor},
  \bibinfo{title}{Zellteilung und Strahlung}, \bibinfo{publisher}{Springer},
  \bibinfo{address}{Berlin}, \bibinfo{year}{1928}.
%Type = Article
\bibitem[{Allibone(1980)}]{Allibone-80}
\bibinfo{author}{T.~E. Allibone}, \bibinfo{journal}{Biograph. Mem. Fell. Roy.
  Soc.} \bibinfo{volume}{26} (\bibinfo{year}{1980}) \bibinfo{pages}{106--47}.
%Type = Book
\bibitem[{Biedenharn and Louck(1981)}]{BL}
\bibinfo{author}{L.~Biedenharn}, \bibinfo{author}{J.~Louck},
  \bibinfo{title}{Angular Momentum in Quantum Physics},
  volume~\bibinfo{volume}{8} of \textit{\bibinfo{series}{Encyclopedia of
  Mathematics and its Applications}}, \bibinfo{publisher}{Addison-Wesley},
  \bibinfo{address}{Reading}, \bibinfo{year}{1981}.
%Type = Book
\bibitem[{Born and Wolf(1985)}]{bornWolf1985}
\bibinfo{author}{M.~Born}, \bibinfo{author}{E.~Wolf},
  \bibinfo{title}{Principles of Optics, 6$^{th}$ ed.}, \bibinfo{edition}{sixth}
  ed., \bibinfo{publisher}{Cambridge University Press},
  \bibinfo{address}{Cambridge}, \bibinfo{year}{1985}.
%Type = Article
\bibitem[{Honegger and Rieckers(2001)}]{Honegger-01}
\bibinfo{author}{R.~Honegger}, \bibinfo{author}{A.~Rieckers},
  \bibinfo{journal}{Ann. Phys.} \bibinfo{volume}{289} (\bibinfo{year}{2001})
  \bibinfo{pages}{213--31}.
%Type = Article
\bibitem[{Honegger and Rieckers(2004)}]{Honegger-04}
\bibinfo{author}{R.~Honegger}, \bibinfo{author}{A.~Rieckers},
  \bibinfo{journal}{Physica A} \bibinfo{volume}{335} (\bibinfo{year}{2004})
  \bibinfo{pages}{487--510}.
%Type = Article
\bibitem[{Skoda(2007)}]{Skoda-07}
\bibinfo{author}{Z.~Skoda}, \bibinfo{journal}{Lett. Math. Phys.}
  \bibinfo{volume}{81} (\bibinfo{year}{2007}) \bibinfo{pages}{1--17}.
%Type = Article
\bibitem[{Schr{\"o}dinger(1926)}]{Schrodinger-26-coherent}
\bibinfo{author}{E.~Schr{\"o}dinger}, \bibinfo{journal}{Naturwissenschaften}
  \bibinfo{volume}{14} (\bibinfo{year}{1926}) \bibinfo{pages}{664--6}.
%Type = Article
\bibitem[{Schwinger(1953)}]{Schwinger-53}
\bibinfo{author}{J.~Schwinger}, \bibinfo{journal}{Phys. Rev.}
  \bibinfo{volume}{91} (\bibinfo{year}{1953}) \bibinfo{pages}{728--40}.
%Type = Article
\bibitem[{Glauber(1963)}]{Glauber-63}
\bibinfo{author}{R.~J. Glauber}, \bibinfo{journal}{Phys. Rev.}
  \bibinfo{volume}{131} (\bibinfo{year}{1963}) \bibinfo{pages}{2766--88}.
%Type = Book
\bibitem[{Itzykson and Zuber(1980)}]{Itzykson}
\bibinfo{author}{C.~Itzykson}, \bibinfo{author}{J.-B. Zuber},
  \bibinfo{title}{Quantum Field Theory}, \bibinfo{publisher}{McGraw-Hill},
  \bibinfo{address}{New York}, \bibinfo{year}{1980}.
%Type = Article
\bibitem[{Pegg and Jeffers(2005)}]{Pegg-05}
\bibinfo{author}{D.~T. Pegg}, \bibinfo{author}{J.~Jeffers},
  \bibinfo{journal}{J. Mod. Optics} \bibinfo{volume}{52} (\bibinfo{year}{2005})
  \bibinfo{pages}{1835--56}.
%Type = Article
\bibitem[{Pegg(2009)}]{Pegg-09}
\bibinfo{author}{D.~T. Pegg}, \bibinfo{journal}{Phys. Rev. A}
  \bibinfo{volume}{79} (\bibinfo{year}{2009}) \bibinfo{pages}{053837}.
%Type = Article
\bibitem[{Loudon and Knight(1987)}]{loudon1987}
\bibinfo{author}{R.~Loudon}, \bibinfo{author}{P.~L. Knight},
  \bibinfo{journal}{Journal of Modern Optics} \bibinfo{volume}{34}
  (\bibinfo{year}{1987}) \bibinfo{pages}{709--759}.
%Type = Article
\bibitem[{Teich(1989)}]{teich1989}
\bibinfo{author}{M.~Teich}, \bibinfo{journal}{Biomedical Engineering, IEEE
  Transactions on} \bibinfo{volume}{36} (\bibinfo{year}{1989})
  \bibinfo{pages}{150--160}.
%Type = Article
\bibitem[{Yuen(1976)}]{Yuen-76}
\bibinfo{author}{H.~P. Yuen}, \bibinfo{journal}{Phys. Rev. A}
  \bibinfo{volume}{13} (\bibinfo{year}{1976}) \bibinfo{pages}{2226--43}.
%Type = Book
\bibitem[{Orszag(2008)}]{orszag2008}
\bibinfo{author}{M.~Orszag}, \bibinfo{title}{Quantum optics: including noise
  reduction, trapped ions, quantum trajectories, and decoherence},
  \bibinfo{publisher}{Springer}, \bibinfo{year}{2008}.
%Type = Article
\bibitem[{Caves(1981)}]{caves1981}
\bibinfo{author}{C.~M. Caves}, \bibinfo{journal}{Phys. Rev. D}
  \bibinfo{volume}{23} (\bibinfo{year}{1981}) \bibinfo{pages}{1693--1708}.
  \URLprefix \url{http://link.aps.org/doi/10.1103/PhysRevD.23.1693}.
  \DOIprefix\doi{10.1103/PhysRevD.23.1693}.
%Type = Book
\bibitem[{Walls and Milburn(2008)}]{walls2008}
\bibinfo{author}{D.~F. Walls}, \bibinfo{author}{G.~J. Milburn},
  \bibinfo{title}{Quantum optics}, \bibinfo{publisher}{Springer},
  \bibinfo{year}{2008}.
%Type = Incollection
\bibitem[{Yuen(2004)}]{yuen2004}
\bibinfo{author}{H.~P. Yuen}, in: \bibinfo{booktitle}{Quantum Squeezing},
  \bibinfo{publisher}{Springer}, \bibinfo{year}{2004}, pp.
  \bibinfo{pages}{227--261}.
%Type = Article
\bibitem[{Saleh and Teich(1987)}]{saleh1987}
\bibinfo{author}{B.~Saleh}, \bibinfo{author}{M.~Teich},
  \bibinfo{journal}{Physical Review Letters} \bibinfo{volume}{58}
  (\bibinfo{year}{1987}) \bibinfo{pages}{2656}.
%Type = Article
\bibitem[{Jiang et~al.(2003)Jiang, Jedrkiewicz, Minardi, Trapani, Mosset,
  Lantz, and Devaux}]{Jiang-03-CCD}
\bibinfo{author}{Y.~Jiang}, \bibinfo{author}{O.~Jedrkiewicz},
  \bibinfo{author}{S.~Minardi}, \bibinfo{author}{P.~D. Trapani},
  \bibinfo{author}{A.~Mosset}, \bibinfo{author}{E.~Lantz},
  \bibinfo{author}{F.~Devaux}, \bibinfo{journal}{Euro. Phys. J. D}
  \bibinfo{volume}{22} (\bibinfo{year}{2003}) \bibinfo{pages}{521--6}.
%Type = Book
\bibitem[{Mosset(2004)}]{Mosset-04-PhD}
\bibinfo{author}{A.~Mosset}, \bibinfo{title}{Etude exp{\'e}rimentale des
  fluctuations spatiales d'origine quantique en amplification param{\'e}trique
  d'images}, \bibinfo{publisher}{Ph.D. thesis}, \bibinfo{address}{University of
  Franche Comt{\'e}}, \bibinfo{year}{2004}.
%Type = Article
\bibitem[{Perina(1967)}]{Perina-67}
\bibinfo{author}{J.~Perina}, \bibinfo{journal}{Phys. Lett. A}
  \bibinfo{volume}{24} (\bibinfo{year}{1967}) \bibinfo{pages}{333--4}.
%Type = Article
\bibitem[{Perina(2010)}]{Perina-10}
\bibinfo{author}{J.~Perina}, \bibinfo{journal}{J. Euro. Opt. Soc.}
  \bibinfo{volume}{5} (\bibinfo{year}{2010}) \bibinfo{pages}{10048s}.
%Type = Article
\bibitem[{Dicke(1954)}]{Dicke-54}
\bibinfo{author}{R.~H. Dicke}, \bibinfo{journal}{Phys. Rev.}
  \bibinfo{volume}{93} (\bibinfo{year}{1954}) \bibinfo{pages}{99--110}.
%Type = Article
\bibitem[{Hepp and Lieb(1973)}]{Hepp-73}
\bibinfo{author}{K.~Hepp}, \bibinfo{author}{E.~H. Lieb},
  \bibinfo{journal}{Annals Phys.} \bibinfo{volume}{76} (\bibinfo{year}{1973})
  \bibinfo{pages}{360--404}.
%Type = Article
\bibitem[{Gross and Haroche(1982)}]{gross1982}
\bibinfo{author}{M.~Gross}, \bibinfo{author}{S.~Haroche},
  \bibinfo{journal}{Physics Reports} \bibinfo{volume}{93}
  (\bibinfo{year}{1982}) \bibinfo{pages}{301--396}.
%Type = Book
\bibitem[{Benedict et~al.(1996)Benedict, Ermolaev, Malyshev, Sokolov, and
  Trifonov}]{Benedict-96}
\bibinfo{author}{M.~G. Benedict}, \bibinfo{author}{A.~M. Ermolaev},
  \bibinfo{author}{V.~A. Malyshev}, \bibinfo{author}{I.~V. Sokolov},
  \bibinfo{author}{E.~D. Trifonov}, \bibinfo{title}{Super-radiance Multiatomic
  Coherent Emission}, \bibinfo{publisher}{Institute of Physics Publishing},
  \bibinfo{address}{Bristol}, \bibinfo{year}{1996}.
%Type = Article
\bibitem[{Emary and Brandes(2003)}]{Emary-03}
\bibinfo{author}{C.~Emary}, \bibinfo{author}{T.~Brandes},
  \bibinfo{journal}{Phys. Rev. E} \bibinfo{volume}{67} (\bibinfo{year}{2003})
  \bibinfo{pages}{066203}.
%Type = Inproceedings
\bibitem[{Trifonov(1999)}]{Trifonov-99}
\bibinfo{author}{E.~D. Trifonov}, in: \bibinfo{editor}{V.~V. Tuchin},
  \bibinfo{editor}{V.~P. Ryubakho}, \bibinfo{editor}{D.~A. Zimnyakov} (Eds.),
  \bibinfo{booktitle}{Light Scattering Technologies for Mechanics, Biomedecine,
  and Material Science}, volume \bibinfo{volume}{3726} of
  \textit{\bibinfo{series}{Proceedings of SPIE}}, \bibinfo{publisher}{SPIE},
  \bibinfo{year}{1999}, pp. \bibinfo{pages}{125--37}.
%Type = Article
\bibitem[{Nagashima(1978)}]{Nagashima-78}
\bibinfo{author}{M.~Nagashima}, \bibinfo{journal}{J. Phys. Soc. Japan}
  \bibinfo{volume}{44} (\bibinfo{year}{1978}) \bibinfo{pages}{1647--55}.
%Type = Article
\bibitem[{Bonifacio and Preparata(1970)}]{Bonifacio-70}
\bibinfo{author}{R.~Bonifacio}, \bibinfo{author}{G.~Preparata},
  \bibinfo{journal}{Phys. Rev. A} \bibinfo{volume}{2} (\bibinfo{year}{1970})
  \bibinfo{pages}{336--47}.
%Type = Article
\bibitem[{Lorenz(1934)}]{Lorenz-34}
\bibinfo{author}{E.~Lorenz}, \bibinfo{journal}{J. Gen. Physiol.}
  \bibinfo{volume}{17} (\bibinfo{year}{1934}) \bibinfo{pages}{843--62}.
%Type = Article
\bibitem[{Gray and Ouellet(1933)}]{Gray-33}
\bibinfo{author}{J.~Gray}, \bibinfo{author}{C.~Ouellet},
  \bibinfo{journal}{Proc. Roy. Soc. Lond. B} \bibinfo{volume}{114}
  (\bibinfo{year}{1933}) \bibinfo{pages}{1--9}.
%Type = Article
\bibitem[{Strehler and Arnold(1951)}]{Strehler-51}
\bibinfo{author}{B.~L. Strehler}, \bibinfo{author}{W.~Arnold},
  \bibinfo{journal}{J. Gen. Physiol.} \bibinfo{volume}{34}
  (\bibinfo{year}{1951}) \bibinfo{pages}{809--20}.
%Type = Article
\bibitem[{Colli and Facchini(1954)}]{colli1}
\bibinfo{author}{L.~Colli}, \bibinfo{author}{U.~Facchini}, \bibinfo{journal}{Il
  Nuovo Cimento} \bibinfo{volume}{12} (\bibinfo{year}{1954})
  \bibinfo{pages}{150--153}.
%Type = Book
\bibitem[{Ruth(1977)}]{Ruth-77}
\bibinfo{author}{B.~Ruth}, \bibinfo{title}{Experimenteller Nachweis
  ultraschwacher Photonenemission aus biologischen Systemen},
  \bibinfo{publisher}{Ph.D. thesis}, \bibinfo{address}{University of Marburg},
  \bibinfo{year}{1977}.
%Type = Article
\bibitem[{Kobayashi and Inaba(2000)}]{Kobayashi-00}
\bibinfo{author}{M.~Kobayashi}, \bibinfo{author}{H.~Inaba},
  \bibinfo{journal}{Applied Optics} \bibinfo{volume}{39} (\bibinfo{year}{2000})
  \bibinfo{pages}{183--92}.
%Type = Article
\bibitem[{Kobayashi(2003{\natexlab{a}})}]{Kobayashi-03}
\bibinfo{author}{M.~Kobayashi}, \bibinfo{journal}{Trends Photochem. Photobiol.}
  \bibinfo{volume}{10} (\bibinfo{year}{2003}{\natexlab{a}})
  \bibinfo{pages}{111--35}.
%Type = Inproceedings
\bibitem[{Kobayashi(2003{\natexlab{b}})}]{Kobayashi-03-conf}
\bibinfo{author}{M.~Kobayashi}, in: \bibinfo{editor}{F.~Musumeci},
  \bibinfo{editor}{L.~S. Brizhik}, \bibinfo{editor}{M.~Ho} (Eds.),
  \bibinfo{booktitle}{Energy and Information Transfer in Biological Systems},
  \bibinfo{publisher}{World Scientific}, \bibinfo{address}{Singapore},
  \bibinfo{year}{2003}{\natexlab{b}}, pp. \bibinfo{pages}{157--87}.
%Type = Inproceedings
\bibitem[{Shen(2003)}]{Shen-03}
\bibinfo{author}{X.~Shen}, in: \bibinfo{editor}{F.-A. Popp},
  \bibinfo{editor}{L.~V. Beloussov} (Eds.), \bibinfo{booktitle}{Integrative
  Biophysics -- Biophotonics}, \bibinfo{publisher}{Kluwer Academic Publishers},
  \bibinfo{address}{Dordrecht}, \bibinfo{year}{2003}, pp.
  \bibinfo{pages}{287--305}.
%Type = Book
\bibitem[{Schirmacher(2008)}]{SchirmacherPhD}
\bibinfo{author}{E.~Schirmacher}, \bibinfo{title}{Untersuchung des {L}ichts
  biologischer {P}roben hinsichtlich nichtklassiker {Z}ust{\"a}nde},
  \bibinfo{publisher}{Ph.D. thesis}, \bibinfo{address}{University of Mainz},
  \bibinfo{year}{2008}.
%Type = Article
\bibitem[{Konak et~al.(1982)Konak, Stepanek, Dvorak, Kupka, Krepelka, and
  Perina}]{Konak-82}
\bibinfo{author}{C.~Konak}, \bibinfo{author}{P.~Stepanek},
  \bibinfo{author}{L.~Dvorak}, \bibinfo{author}{Z.~Kupka},
  \bibinfo{author}{J.~Krepelka}, \bibinfo{author}{J.~Perina},
  \bibinfo{journal}{Optica Acta} \bibinfo{volume}{29} (\bibinfo{year}{1982})
  \bibinfo{pages}{1105--16}.
%Type = Article
\bibitem[{Hill et~al.(1998)Hill, Barnes, Lermer, Whitten, and Ramsey}]{Hill-98}
\bibinfo{author}{S.~C. Hill}, \bibinfo{author}{M.~D. Barnes},
  \bibinfo{author}{N.~Lermer}, \bibinfo{author}{W.~B. Whitten},
  \bibinfo{author}{J.~M. Ramsey}, \bibinfo{journal}{Anal. Chem.}
  \bibinfo{volume}{70} (\bibinfo{year}{1998}) \bibinfo{pages}{2964--71}.
%Type = Article
\bibitem[{Kobayashi et~al.(1998)Kobayashi, Deveraj, and Inaba}]{Kobayashi-98}
\bibinfo{author}{M.~Kobayashi}, \bibinfo{author}{B.~Deveraj},
  \bibinfo{author}{H.~Inaba}, \bibinfo{journal}{Phys. Rev. E}
  \bibinfo{volume}{57} (\bibinfo{year}{1998}) \bibinfo{pages}{2129--33}.
%Type = Article
\bibitem[{Huang et~al.(2005)Huang, Shao, Wang, Xiao, and Tia}]{Huang-05}
\bibinfo{author}{T.~Huang}, \bibinfo{author}{J.~Shao},
  \bibinfo{author}{X.~Wang}, \bibinfo{author}{L.~Xiao},
  \bibinfo{author}{S.~Tia}, \bibinfo{journal}{Opt. Eng.} \bibinfo{volume}{44}
  (\bibinfo{year}{2005}) \bibinfo{pages}{074001}.
%Type = Article
\bibitem[{Collinson and Wightman(1995)}]{Collinson-95}
\bibinfo{author}{M.~M. Collinson}, \bibinfo{author}{R.~M. Wightman},
  \bibinfo{journal}{Science} \bibinfo{volume}{268} (\bibinfo{year}{1995})
  \bibinfo{pages}{1883--5}.
%Type = Article
\bibitem[{Lindner(2006)}]{Lindner-06}
\bibinfo{author}{B.~Lindner}, \bibinfo{journal}{Phys. Rev. E}
  \bibinfo{volume}{73} (\bibinfo{year}{2006}) \bibinfo{pages}{022901}.
%Type = Article
\bibitem[{C{\^a}teau and Reyes(2006)}]{Cateau-06}
\bibinfo{author}{H.~C{\^a}teau}, \bibinfo{author}{A.~D. Reyes},
  \bibinfo{journal}{Phys. Rev. Lett.} \bibinfo{volume}{96}
  (\bibinfo{year}{2006}) \bibinfo{pages}{058101}.
%Type = Article
\bibitem[{Banys(1977)}]{Banys-77}
\bibinfo{author}{R.~Banys}, \bibinfo{journal}{Lithuanian Math. J.}
  \bibinfo{volume}{17} (\bibinfo{year}{1977}) \bibinfo{pages}{11--6}.
%Type = Article
\bibitem[{Saleh et~al.(1983)Saleh, Stoler, and Teich}]{Saleh-83}
\bibinfo{author}{B.~E.~A. Saleh}, \bibinfo{author}{D.~Stoler},
  \bibinfo{author}{M.~C. Teich}, \bibinfo{journal}{Phys. Rev. A}
  \bibinfo{volume}{27} (\bibinfo{year}{1983}) \bibinfo{pages}{360--74}.
%Type = Article
\bibitem[{Kobayashi et~al.(1999)Kobayashi, Takeda, Ito, Kato, and
  Inaba}]{Kobayashi-99}
\bibinfo{author}{M.~Kobayashi}, \bibinfo{author}{M.~Takeda},
  \bibinfo{author}{K.-I. Ito}, \bibinfo{author}{H.~Kato},
  \bibinfo{author}{H.~Inaba}, \bibinfo{journal}{J. Neurosci. Meth.}
  \bibinfo{volume}{93} (\bibinfo{year}{1999}) \bibinfo{pages}{163--8}.
%Type = Inproceedings
\bibitem[{Kobayashi(2005)}]{Kobayashi-05}
\bibinfo{author}{M.~Kobayashi}, in: \bibinfo{editor}{X.~Shen},
  \bibinfo{editor}{R.~van Wijk} (Eds.), \bibinfo{booktitle}{Biophotonics --
  Optical Science and Engineering for the 21st Century},
  \bibinfo{publisher}{Springer}, \bibinfo{address}{New York},
  \bibinfo{year}{2005}, pp. \bibinfo{pages}{155--71}.
%Type = Article
\bibitem[{Scott et~al.(1989)Scott, Usa, and Inaba}]{Scott-89}
\bibinfo{author}{R.~Q. Scott}, \bibinfo{author}{M.~Usa},
  \bibinfo{author}{H.~Inaba}, \bibinfo{journal}{Appl. Phys. B}
  \bibinfo{volume}{48} (\bibinfo{year}{1989}) \bibinfo{pages}{183--5}.
%Type = Inproceedings
\bibitem[{Inaba(1990)}]{Inaba-90}
\bibinfo{author}{H.~Inaba}, in: \bibinfo{editor}{J.~B. Andersen} (Ed.),
  \bibinfo{booktitle}{Modern Radio Science 1990}, \bibinfo{publisher}{Oxford
  University Press}, \bibinfo{address}{Oxford}, \bibinfo{year}{1990}, pp.
  \bibinfo{pages}{163--84}.
%Type = Article
\bibitem[{Kobayashi et~al.(1996)Kobayashi, Deveraj, Usa, Tanno, Takeda, and
  Inaba}]{Kobayashi-96}
\bibinfo{author}{M.~Kobayashi}, \bibinfo{author}{B.~Deveraj},
  \bibinfo{author}{M.~Usa}, \bibinfo{author}{Y.~Tanno},
  \bibinfo{author}{M.~Takeda}, \bibinfo{author}{H.~Inaba},
  \bibinfo{journal}{Frontiers of Medical and Biological Engineering}
  \bibinfo{volume}{7} (\bibinfo{year}{1996}) \bibinfo{pages}{299--309}.
%Type = Article
\bibitem[{Kobayashi et~al.(1997)Kobayashi, Deveraj, Usa, Tanno, Takedan, and
  Inaba}]{Kobayashi-97}
\bibinfo{author}{M.~Kobayashi}, \bibinfo{author}{B.~Deveraj},
  \bibinfo{author}{M.~Usa}, \bibinfo{author}{Y.~Tanno},
  \bibinfo{author}{M.~Takedan}, \bibinfo{author}{H.~Inaba},
  \bibinfo{journal}{Photochem. Photobiol.} \bibinfo{volume}{65}
  (\bibinfo{year}{1997}) \bibinfo{pages}{535--7}.
%Type = Article
\bibitem[{Williams et~al.(1982)Williams, Jr, and Chance}]{Williams-82}
\bibinfo{author}{M.~D. Williams}, \bibinfo{author}{J.~S.~L. Jr},
  \bibinfo{author}{B.~Chance}, \bibinfo{journal}{Ann. New York Acad. Sci.}
  \bibinfo{volume}{386} (\bibinfo{year}{1982}) \bibinfo{pages}{478--83}.
%Type = Inproceedings
\bibitem[{Slawinska and Slawinski(1985)}]{Slawinska}
\bibinfo{author}{D.~Slawinska}, \bibinfo{author}{J.~Slawinski}, in:
  \bibinfo{editor}{J.~G. Burr} (Ed.), \bibinfo{booktitle}{Chemi- and
  Bioluminescence}, \bibinfo{publisher}{Marcel Dekker}, \bibinfo{address}{New
  York}, \bibinfo{year}{1985}, pp. \bibinfo{pages}{495--531}.
%Type = Article
\bibitem[{Shen et~al.(1993)Shen, Liu, and Li}]{Shen-93}
\bibinfo{author}{X.~Shen}, \bibinfo{author}{F.~Liu}, \bibinfo{author}{X.~Y.
  Li}, \bibinfo{journal}{Experientia} \bibinfo{volume}{49}
  (\bibinfo{year}{1993}) \bibinfo{pages}{291--5}.
%Type = Inproceedings
\bibitem[{Gallep et~al.(2004)Gallep, Conforti, Braghini, Maluf, Yan, and
  Popp}]{Gallep-04}
\bibinfo{author}{C.~M. Gallep}, \bibinfo{author}{E.~Conforti},
  \bibinfo{author}{M.~T. Braghini}, \bibinfo{author}{M.~P. Maluf},
  \bibinfo{author}{Y.~Yan}, \bibinfo{author}{F.-A. Popp}, in:
  \bibinfo{booktitle}{11th Brazilian Symposium on Microwaves and
  Optoelectronics, S{\~{a}}o Paulo}.
%Type = Inproceedings
\bibitem[{Gallep et~al.(2007)Gallep, Batista, Pereira, Oliveira, and
  Siqueira}]{Gallep-07}
\bibinfo{author}{C.~M. Gallep}, \bibinfo{author}{D.~C. Batista},
  \bibinfo{author}{C.~A. Pereira}, \bibinfo{author}{V.~M. Oliveira},
  \bibinfo{author}{N.~A. Siqueira}, in: \bibinfo{booktitle}{Proceedings of 2007
  SBMO/IEEE MTT-S International Microwave and Optoelectronics Conference,
  Salvador, Brazil}, pp. \bibinfo{pages}{241--4}.
%Type = Article
\bibitem[{van Wijk et~al.(2006)van Wijk, van Wijk, and Bajpai}]{Wijk-06}
\bibinfo{author}{R.~van Wijk}, \bibinfo{author}{E.~P.~A. van Wijk},
  \bibinfo{author}{R.~P. Bajpai}, \bibinfo{journal}{J. Photochem. Photobiol. B}
  \bibinfo{volume}{84} (\bibinfo{year}{2006}) \bibinfo{pages}{46--55}.
%Type = Article
\bibitem[{van Wijk et~al.(2010)van Wijk, van Wijk, Bajpai, and van~der
  Greef}]{Wijk-10}
\bibinfo{author}{E.~P.~A. van Wijk}, \bibinfo{author}{R.~van Wijk},
  \bibinfo{author}{R.~P. Bajpai}, \bibinfo{author}{J.~van~der Greef},
  \bibinfo{journal}{J. Photochem. Photobiol. B} \bibinfo{volume}{99}
  (\bibinfo{year}{2010}) \bibinfo{pages}{133--43}.
%Type = Article
\bibitem[{Bajpai et~al.(2013)Bajpai, Van~Wijk, Van~Wijk, and van~der
  Greef}]{bajpai2013}
\bibinfo{author}{R.~P. Bajpai}, \bibinfo{author}{E.~Van~Wijk},
  \bibinfo{author}{R.~Van~Wijk}, \bibinfo{author}{J.~van~der Greef},
  \bibinfo{journal}{Journal of Photochemistry and Photobiology B: Biology}
  \bibinfo{volume}{129} (\bibinfo{year}{2013}) \bibinfo{pages}{6--16}.
%Type = Article
\bibitem[{Ruth and Popp(1976)}]{Ruth-76}
\bibinfo{author}{B.~Ruth}, \bibinfo{author}{F.-A. Popp}, \bibinfo{journal}{Z.
  Natur. C} \bibinfo{volume}{31} (\bibinfo{year}{1976})
  \bibinfo{pages}{741--5}.
%Type = Article
\bibitem[{Hideg et~al.(1990)Hideg, Kobayashi, and Inaba}]{Hideg-90}
\bibinfo{author}{E.~Hideg}, \bibinfo{author}{M.~Kobayashi},
  \bibinfo{author}{H.~Inaba}, \bibinfo{journal}{FEBS Letters}
  \bibinfo{volume}{275} (\bibinfo{year}{1990}) \bibinfo{pages}{121--4}.
%Type = Article
\bibitem[{Schmidt(1987)}]{Schmidt-87}
\bibinfo{author}{W.~Schmidt}, \bibinfo{journal}{Photochem. Photobiol.}
  \bibinfo{volume}{45} (\bibinfo{year}{1987}) \bibinfo{pages}{555--6}.
%Type = Article
\bibitem[{Popp and Ruth(1977)}]{Popp-77}
\bibinfo{author}{F.-A. Popp}, \bibinfo{author}{B.~Ruth},
  \bibinfo{journal}{Arzneimittel-Forschung / Drug Research}
  \bibinfo{volume}{27} (\bibinfo{year}{1977}) \bibinfo{pages}{933--9}.
%Type = Book
\bibitem[{Popp(1976)}]{Popp-76}
\bibinfo{author}{F.-A. Popp}, \bibinfo{title}{Biophotonen. {E}in neuer {W}eg
  zur {L}{\"o}sung des {K}rebsproblems}, \bibinfo{publisher}{Verlag f. Medizin
  Ewald Fischer}, \bibinfo{address}{Heidelberg}, \bibinfo{year}{1976}.
%Type = Book
\bibitem[{Beloussov et~al.(2000)Beloussov, Popp, Voeikov, and
  Wijk}]{bpe-book2000}
\bibinfo{editor}{L.~Beloussov}, \bibinfo{editor}{F.-A. Popp},
  \bibinfo{editor}{V.~Voeikov}, \bibinfo{editor}{R.~V. Wijk} (Eds.),
  \bibinfo{title}{Biophotonics and Coherent Systems},
  \bibinfo{publisher}{Moscow University Press}, \bibinfo{year}{2000}.
%Type = Article
\bibitem[{Fr\"{o}hlich(1977)}]{froh6}
\bibinfo{author}{H.~Fr\"{o}hlich}, \bibinfo{journal}{Rivista del Nuovo Cimento}
  \bibinfo{volume}{7} (\bibinfo{year}{1977}) \bibinfo{pages}{399--418}.
%Type = Article
\bibitem[{Fr\"{o}hlich(1968)}]{froh2}
\bibinfo{author}{H.~Fr\"{o}hlich}, \bibinfo{journal}{International Journal of
  Quantum Chemistry} \bibinfo{volume}{2} (\bibinfo{year}{1968})
  \bibinfo{pages}{641--649}.
%Type = Article
\bibitem[{Fr\"{o}hlich(1970)}]{froh9}
\bibinfo{author}{H.~Fr\"{o}hlich}, \bibinfo{journal}{Nature}
  \bibinfo{volume}{228} (\bibinfo{year}{1970}) \bibinfo{pages}{1093}.
%Type = Article
\bibitem[{Fr\"{o}hlich(1978)}]{froh10}
\bibinfo{author}{H.~Fr\"{o}hlich}, \bibinfo{journal}{IEEE Transactions on
  Microwave Theory and Techniques} \bibinfo{volume}{26} (\bibinfo{year}{1978})
  \bibinfo{pages}{613--618}.
%Type = Book
\bibitem[{Popp et~al.(1992)Popp, Li, and Gu}]{bpe-book1992}
\bibinfo{editor}{F.-A. Popp}, \bibinfo{editor}{K.~H. Li},
  \bibinfo{editor}{Q.~Gu} (Eds.), \bibinfo{title}{Recent Advances in Biophoton
  research and its Applications}, \bibinfo{publisher}{World Scientific:
  Singapure, London, New York, Hong Kong}, \bibinfo{year}{1992}.
%Type = Article
\bibitem[{Popp et~al.(1981)Popp, Ruth, Bahr, B{\"o}hm, Gra{\ss}, Grolig,
  Rattemeyer, Schmidt, and Wulle}]{Popp-81}
\bibinfo{author}{F.-A. Popp}, \bibinfo{author}{B.~Ruth},
  \bibinfo{author}{W.~Bahr}, \bibinfo{author}{J.~B{\"o}hm},
  \bibinfo{author}{P.~Gra{\ss}}, \bibinfo{author}{G.~Grolig},
  \bibinfo{author}{M.~Rattemeyer}, \bibinfo{author}{H.~G. Schmidt},
  \bibinfo{author}{P.~Wulle}, \bibinfo{journal}{Collective Phenomena}
  \bibinfo{volume}{3} (\bibinfo{year}{1981}) \bibinfo{pages}{187--213}.
%Type = Inproceedings
\bibitem[{Popp(1981)}]{Popp-81-wieder}
\bibinfo{author}{F.-A. Popp}, in: \bibinfo{editor}{K.~Theurer},
  \bibinfo{editor}{G.~F. Domagk}, \bibinfo{editor}{H.~Kraft} (Eds.),
  \bibinfo{booktitle}{Wiederherstellung und Erneuerung als Prinzipien der
  Organo- und Immunotherapie}, \bibinfo{publisher}{Enke},
  \bibinfo{address}{Stuttgart}, \bibinfo{year}{1981}, pp.
  \bibinfo{pages}{48--67}.
%Type = Article
\bibitem[{Popp(1983)}]{Popp-83-Arzt}
\bibinfo{author}{F.-A. Popp}, \bibinfo{journal}{{\"A}rztezeitschrift f{\"u}r
  {N}aturheilverfahren} \bibinfo{volume}{24} (\bibinfo{year}{1983})
  \bibinfo{pages}{361--6}.
%Type = Article
\bibitem[{Popp et~al.(1984)Popp, Nagl, Li, Scholz, Weing{\"a}rtner, and
  Wolf}]{Popp-84}
\bibinfo{author}{F.-A. Popp}, \bibinfo{author}{W.~Nagl}, \bibinfo{author}{K.~H.
  Li}, \bibinfo{author}{W.~Scholz}, \bibinfo{author}{O.~Weing{\"a}rtner},
  \bibinfo{author}{R.~Wolf}, \bibinfo{journal}{Cell Biophys.}
  \bibinfo{volume}{6} (\bibinfo{year}{1984}) \bibinfo{pages}{33--52}.
%Type = Inproceedings
\bibitem[{Popp(1986)}]{Popp-86-Pasteur}
\bibinfo{author}{F.-A. Popp}, in: \bibinfo{booktitle}{Louis Pasteur, Source
  d'une Nouvelle Renaisance}, \bibinfo{publisher}{Fondation pour l'Energie de
  la Fusion}, \bibinfo{address}{Paris}, \bibinfo{year}{1986}, pp.
  \bibinfo{pages}{169--77}.
%Type = Inproceedings
\bibitem[{Popp(1995)}]{Popp-95}
\bibinfo{author}{F.-A. Popp}, in: \bibinfo{editor}{L.~V. Beloussov},
  \bibinfo{editor}{F.-A. Popp} (Eds.), \bibinfo{booktitle}{Biophotonics --
  Non-equilibrium and Coherent Systems in Biology, Biophysics and
  Biotechnology}, \bibinfo{publisher}{Bioinform Services Co},
  \bibinfo{address}{Moscow}, \bibinfo{year}{1995}, pp. \bibinfo{pages}{85--98}.
%Type = Inproceedings
\bibitem[{Li et~al.(1983)Li, Popp, Nagl, and Klima}]{Li-83}
\bibinfo{author}{K.~H. Li}, \bibinfo{author}{F.-A. Popp},
  \bibinfo{author}{W.~Nagl}, \bibinfo{author}{H.~Klima}, in:
  \bibinfo{editor}{H.~Fr{\"o}hlich}, \bibinfo{editor}{F.~Kremer} (Eds.),
  \bibinfo{booktitle}{Coherent Excitations in Biological Systems},
  \bibinfo{publisher}{Springer}, \bibinfo{address}{Berlin},
  \bibinfo{year}{1983}, pp. \bibinfo{pages}{117--27}.
%Type = Article
\bibitem[{Salari and Brouder(2011)}]{Salari-11}
\bibinfo{author}{V.~Salari}, \bibinfo{author}{C.~Brouder},
  \bibinfo{journal}{Phys. Lett. A} \bibinfo{volume}{375} (\bibinfo{year}{2011})
  \bibinfo{pages}{2531--2}.
%Type = Article
\bibitem[{Goltsev et~al.(2009)Goltsev, Zaharieva, Chernev, and
  Strasser}]{Goltsev-09}
\bibinfo{author}{V.~Goltsev}, \bibinfo{author}{I.~Zaharieva},
  \bibinfo{author}{P.~Chernev}, \bibinfo{author}{R.~J. Strasser},
  \bibinfo{journal}{Photosynthesis Research} \bibinfo{volume}{101}
  (\bibinfo{year}{2009}) \bibinfo{pages}{217--32}.
%Type = Article
\bibitem[{Guo and Tan(2009)}]{Guo-09}
\bibinfo{author}{Y.~Guo}, \bibinfo{author}{J.-L. Tan},
  \bibinfo{journal}{BioSystems} \bibinfo{volume}{95} (\bibinfo{year}{2009})
  \bibinfo{pages}{98--103}.
%Type = Inproceedings
\bibitem[{Popp and Li(1992)}]{Popp-Li-92}
\bibinfo{author}{F.-A. Popp}, \bibinfo{author}{K.~H. Li}, in:
  \bibinfo{editor}{F.-A. Popp}, \bibinfo{editor}{K.~H. Li},
  \bibinfo{editor}{Q.~Gu} (Eds.), \bibinfo{booktitle}{Recent Advances in
  Biophoton Research and its Applications}, \bibinfo{publisher}{World
  Scientific}, \bibinfo{address}{Singapore}, \bibinfo{year}{1992}, pp.
  \bibinfo{pages}{47--58}.
%Type = Article
\bibitem[{Popp and Li(1993)}]{Popp-93}
\bibinfo{author}{F.-A. Popp}, \bibinfo{author}{K.-H. Li},
  \bibinfo{journal}{Int. J. Theor. Phys.} \bibinfo{volume}{32}
  (\bibinfo{year}{1993}) \bibinfo{pages}{1573--83}.
%Type = Article
\bibitem[{Popp and Yan(2002)}]{Popp-02}
\bibinfo{author}{F.-A. Popp}, \bibinfo{author}{Y.~Yan}, \bibinfo{journal}{Phys.
  Lett. A} \bibinfo{volume}{293} (\bibinfo{year}{2002}) \bibinfo{pages}{93--7}.
%Type = Article
\bibitem[{Yan et~al.(2005)Yan, Popp, Sigrist, Schlesinger, Dolf, Yan, Cohen,
  and Chotia}]{Yan-05}
\bibinfo{author}{Y.~Yan}, \bibinfo{author}{F.-A. Popp},
  \bibinfo{author}{S.~Sigrist}, \bibinfo{author}{D.~Schlesinger},
  \bibinfo{author}{A.~Dolf}, \bibinfo{author}{Z.~Yan},
  \bibinfo{author}{S.~Cohen}, \bibinfo{author}{A.~Chotia}, \bibinfo{journal}{J.
  Photochem. Photobiol. B} \bibinfo{volume}{78} (\bibinfo{year}{2005})
  \bibinfo{pages}{235--44}.
%Type = Article
\bibitem[{Popp et~al.(1988)Popp, Li, Mei, Galle, and Neurohr}]{Popp-88}
\bibinfo{author}{F.-A. Popp}, \bibinfo{author}{K.~H. Li},
  \bibinfo{author}{W.~P. Mei}, \bibinfo{author}{M.~Galle},
  \bibinfo{author}{R.~Neurohr}, \bibinfo{journal}{Experientia}
  \bibinfo{volume}{44} (\bibinfo{year}{1988}) \bibinfo{pages}{576--85}.
%Type = Inproceedings
\bibitem[{Popp(1989)}]{Popp-89}
\bibinfo{author}{F.-A. Popp}, in: \bibinfo{editor}{F.-A. Popp},
  \bibinfo{editor}{U.~Warnke}, \bibinfo{editor}{H.~L. K{\"o}nig},
  \bibinfo{editor}{W.~Peschka} (Eds.), \bibinfo{booktitle}{Electromagnetic
  Bio-Information}, \bibinfo{publisher}{Urban \& Schwartzenberg},
  \bibinfo{address}{Munich}, \bibinfo{year}{1989}, pp.
  \bibinfo{pages}{144--67}.
%Type = Inproceedings
\bibitem[{Popp(1992{\natexlab{a}})}]{Popp-92}
\bibinfo{author}{F.-A. Popp}, in: \bibinfo{editor}{F.-A. Popp},
  \bibinfo{editor}{K.~H. Li}, \bibinfo{editor}{Q.~Gu} (Eds.),
  \bibinfo{booktitle}{Recent Advances in Biophoton Research and its
  Applications}, \bibinfo{publisher}{World Scientific},
  \bibinfo{address}{Singapore}, \bibinfo{year}{1992}{\natexlab{a}}, pp.
  \bibinfo{pages}{1--46}.
%Type = Inproceedings
\bibitem[{Popp(1992{\natexlab{b}})}]{Popp-92-evolution}
\bibinfo{author}{F.-A. Popp}, in: \bibinfo{editor}{F.-A. Popp},
  \bibinfo{editor}{K.~H. Li}, \bibinfo{editor}{Q.~Gu} (Eds.),
  \bibinfo{booktitle}{Recent Advances in Biophoton Research and its
  Applications}, \bibinfo{publisher}{World Scientific},
  \bibinfo{address}{Singapore}, \bibinfo{year}{1992}{\natexlab{b}}, pp.
  \bibinfo{pages}{357--73}.
%Type = Inproceedings
\bibitem[{Li(1992)}]{Li-92}
\bibinfo{author}{K.-H. Li}, in: \bibinfo{editor}{F.-A. Popp},
  \bibinfo{editor}{K.~H. Li}, \bibinfo{editor}{Q.~Gu} (Eds.),
  \bibinfo{booktitle}{Recent Advances in Biophoton Research and its
  Applications}, \bibinfo{publisher}{World Scientific},
  \bibinfo{address}{Singapore}, \bibinfo{year}{1992}, pp.
  \bibinfo{pages}{113--155}.
%Type = Inproceedings
\bibitem[{Chang and Popp(1998)}]{Chang-98}
\bibinfo{author}{J.~J. Chang}, \bibinfo{author}{F.-A. Popp}, in:
  \bibinfo{editor}{J.-J. Chang}, \bibinfo{editor}{J.~Fisch},
  \bibinfo{editor}{F.-A. Popp} (Eds.), \bibinfo{booktitle}{Biophotons},
  \bibinfo{publisher}{Kluwer Academic Publishers},
  \bibinfo{address}{Dordrecht}, \bibinfo{year}{1998}, pp.
  \bibinfo{pages}{217--27}.
%Type = Article
\bibitem[{Popp and Chang(2000)}]{Popp-00}
\bibinfo{author}{F.-A. Popp}, \bibinfo{author}{J.-J. Chang},
  \bibinfo{journal}{Science in China C} \bibinfo{volume}{43}
  (\bibinfo{year}{2000}) \bibinfo{pages}{507--18}.
%Type = Article
\bibitem[{Bajpai et~al.(1998)Bajpai, Kumar, and Sivadasan}]{Bajpai-98-AMC}
\bibinfo{author}{R.~P. Bajpai}, \bibinfo{author}{S.~Kumar},
  \bibinfo{author}{V.~A. Sivadasan}, \bibinfo{journal}{Appl. Math. Comput.}
  \bibinfo{volume}{93} (\bibinfo{year}{1998}) \bibinfo{pages}{277--88}.
%Type = Article
\bibitem[{Chang(2008{\natexlab{a}})}]{Chang-08}
\bibinfo{author}{J.-J. Chang}, \bibinfo{journal}{Indian J. Exp. Biol.}
  \bibinfo{volume}{46} (\bibinfo{year}{2008}{\natexlab{a}})
  \bibinfo{pages}{371--7}.
%Type = Article
\bibitem[{Chang(2008{\natexlab{b}})}]{Chang-08-neuro}
\bibinfo{author}{J.-J. Chang}, \bibinfo{journal}{NeuroQuantology}
  \bibinfo{volume}{6} (\bibinfo{year}{2008}{\natexlab{b}})
  \bibinfo{pages}{420--30}.
%Type = Inproceedings
\bibitem[{Popp(2003)}]{Popp-03}
\bibinfo{author}{F.-A. Popp}, in: \bibinfo{editor}{F.-A. Popp},
  \bibinfo{editor}{L.~V. Beloussov} (Eds.), \bibinfo{booktitle}{Integrative
  Biophysics -- Biophotonics}, \bibinfo{publisher}{Kluwer Academic Publishers},
  \bibinfo{address}{Dordrecht}, \bibinfo{year}{2003}, pp.
  \bibinfo{pages}{387--438}.
%Type = Inproceedings
\bibitem[{Bajpai(2009)}]{Bajpai-09}
\bibinfo{author}{R.~P. Bajpai}, in: \bibinfo{editor}{V.~B. Meyer-Rochow} (Ed.),
  \bibinfo{booktitle}{Bioluminescence in Focus -- A Collection of Illuminating
  Essays}, \bibinfo{publisher}{Research Signpost},
  \bibinfo{address}{Trivandrum}, \bibinfo{year}{2009}, pp.
  \bibinfo{pages}{357--85}.
%Type = Article
\bibitem[{Bajpai(1999)}]{Bajpai-99}
\bibinfo{author}{R.~P. Bajpai}, \bibinfo{journal}{J. Theor. Biol.}
  \bibinfo{volume}{198} (\bibinfo{year}{1999}) \bibinfo{pages}{287--99}.
%Type = Article
\bibitem[{Bajpai(2003)}]{Bajpai-03}
\bibinfo{author}{R.~P. Bajpai}, \bibinfo{journal}{Indian J. Exp. Biol.}
  \bibinfo{volume}{41} (\bibinfo{year}{2003}) \bibinfo{pages}{514--27}.
%Type = Article
\bibitem[{Bajpai(2004)}]{Bajpai-04}
\bibinfo{author}{R.~P. Bajpai}, \bibinfo{journal}{Phys. Lett. A}
  \bibinfo{volume}{322} (\bibinfo{year}{2004}) \bibinfo{pages}{131--6}.
%Type = Inproceedings
\bibitem[{Bajpai(2005)}]{Bajpai-05}
\bibinfo{author}{R.~P. Bajpai}, in: \bibinfo{editor}{X.~Shen},
  \bibinfo{editor}{R.~van Wijk} (Eds.), \bibinfo{booktitle}{Biophotonics --
  Optical Science and Engineering for the 21st Century},
  \bibinfo{publisher}{Springer}, \bibinfo{address}{New York},
  \bibinfo{year}{2005}, pp. \bibinfo{pages}{125--40}.
%Type = Inproceedings
\bibitem[{Zhang(2000)}]{Zhang-00}
\bibinfo{author}{W.~Zhang}, in: \bibinfo{editor}{A.~N. Mira} (Ed.),
  \bibinfo{booktitle}{Quantum Field Theory: A Twentieth Century Profile},
  \bibinfo{publisher}{Hidustan Book Agency}, \bibinfo{address}{Gurgaon},
  \bibinfo{year}{2000}, pp. \bibinfo{pages}{297--323}.
%Type = Article
\bibitem[{Bajpai(2005)}]{Bajpai-05-PLA}
\bibinfo{author}{R.~P. Bajpai}, \bibinfo{journal}{Phys. Lett. A}
  \bibinfo{volume}{337} (\bibinfo{year}{2005}) \bibinfo{pages}{265--73}.
%Type = Article
\bibitem[{Liu and Tombesi(1992)}]{Liu-92}
\bibinfo{author}{W.~S. Liu}, \bibinfo{author}{P.~Tombesi},
  \bibinfo{journal}{Nuovo Cimento B} \bibinfo{volume}{107}
  (\bibinfo{year}{1992}) \bibinfo{pages}{595--602}.
%Type = Inproceedings
\bibitem[{Bajpai(2007)}]{Bajpai-07}
\bibinfo{author}{R.~P. Bajpai}, in: \bibinfo{editor}{L.~V. Beloussov},
  \bibinfo{editor}{V.~L. Voeikov}, \bibinfo{editor}{V.~S. Martynyuk} (Eds.),
  \bibinfo{booktitle}{Biophotonics and Coherent Systems in Biology},
  \bibinfo{publisher}{Springer}, \bibinfo{address}{Berlin},
  \bibinfo{year}{2007}, pp. \bibinfo{pages}{33--46}.
%Type = Article
\bibitem[{Bajpai(2008)}]{Bajpai-08-Xanthoria}
\bibinfo{author}{R.~P. Bajpai}, \bibinfo{journal}{Indian J. Exp. Biol.}
  \bibinfo{volume}{46} (\bibinfo{year}{2008}) \bibinfo{pages}{420--32}.
%Type = Article
\bibitem[{Racine et~al.(2013)Racine, Rastogi, and Bajpai}]{racine2013}
\bibinfo{author}{D.~Racine}, \bibinfo{author}{A.~Rastogi},
  \bibinfo{author}{R.~P. Bajpai}, \bibinfo{journal}{Chinese Medicine}
  \bibinfo{volume}{4} (\bibinfo{year}{2013}) \bibinfo{pages}{72}.
%Type = Article
\bibitem[{Assmann and Bayer(2012)}]{assmann2012}
\bibinfo{author}{M.~Assmann}, \bibinfo{author}{M.~Bayer},
  \bibinfo{journal}{Optics Letters} \bibinfo{volume}{37} (\bibinfo{year}{2012})
  \bibinfo{pages}{2811--2813}.
%Type = Article
\bibitem[{del Giudice et~al.(2005)del Giudice, de~Ninno, Fleischmann, Mengoli,
  Milani, Talpo, and Vitiello}]{Giudice-05}
\bibinfo{author}{E.~del Giudice}, \bibinfo{author}{A.~de~Ninno},
  \bibinfo{author}{M.~Fleischmann}, \bibinfo{author}{G.~Mengoli},
  \bibinfo{author}{M.~Milani}, \bibinfo{author}{G.~Talpo},
  \bibinfo{author}{G.~Vitiello}, \bibinfo{journal}{Electromag. Biol. Med.}
  \bibinfo{volume}{24} (\bibinfo{year}{2005}) \bibinfo{pages}{199--210}.
%Type = Article
\bibitem[{Greffet et~al.(2002)Greffet, Carminati, Joulain, Mulet, Mainguy, and
  Chen}]{Greffet-02}
\bibinfo{author}{J.~J. Greffet}, \bibinfo{author}{R.~Carminati},
  \bibinfo{author}{K.~Joulain}, \bibinfo{author}{J.~P. Mulet},
  \bibinfo{author}{S.~Mainguy}, \bibinfo{author}{Y.~Chen},
  \bibinfo{journal}{Nature} \bibinfo{volume}{416} (\bibinfo{year}{2002})
  \bibinfo{pages}{61--4}.
%Type = Article
\bibitem[{Zhai et~al.(2006)Zhai, Chen, and Wu}]{Zhai-06}
\bibinfo{author}{Y.~H. Zhai}, \bibinfo{author}{X.~H. Chen},
  \bibinfo{author}{L.~A. Wu}, \bibinfo{journal}{Phys. Rev. A}
  \bibinfo{volume}{74} (\bibinfo{year}{2006}) \bibinfo{pages}{053807}.
%Type = Article
\bibitem[{Pittman et~al.(1996)Pittman, Strekalov, Migdall, Rubin, Sergienko,
  and Shih}]{Pittman-96}
\bibinfo{author}{T.~B. Pittman}, \bibinfo{author}{D.~V. Strekalov},
  \bibinfo{author}{A.~Migdall}, \bibinfo{author}{M.~H. Rubin},
  \bibinfo{author}{A.~V. Sergienko}, \bibinfo{author}{Y.~H. Shih},
  \bibinfo{journal}{Phys. Rev. Lett.} \bibinfo{volume}{77}
  (\bibinfo{year}{1996}) \bibinfo{pages}{1917--20}.
%Type = Inproceedings
\bibitem[{Letokhov and Dobryakov(2003)}]{Letokhov-03}
\bibinfo{author}{V.~S. Letokhov}, \bibinfo{author}{A.~L. Dobryakov}, in:
  \bibinfo{editor}{F.~Musumeci}, \bibinfo{editor}{L.~S. Brizhik},
  \bibinfo{editor}{M.~Ho} (Eds.), \bibinfo{booktitle}{Energy and Information
  Transfer in Biological Systems}, \bibinfo{publisher}{World Scientific},
  \bibinfo{address}{Singapore}, \bibinfo{year}{2003}, pp.
  \bibinfo{pages}{205--16}.
%Type = Article
\bibitem[{Parson(2007)}]{parson2007}
\bibinfo{author}{W.~Parson}, \bibinfo{journal}{Science} \bibinfo{volume}{316}
  (\bibinfo{year}{2007}) \bibinfo{pages}{1438--1439}.
%Type = Article
\bibitem[{Wolynes(2009)}]{wolynes2009}
\bibinfo{author}{P.~Wolynes}, \bibinfo{journal}{Proceedings of the National
  Academy of Sciences} \bibinfo{volume}{106} (\bibinfo{year}{2009})
  \bibinfo{pages}{17247}.
%Type = Article
\bibitem[{Engel et~al.(2007)Engel, Calhoun, Read, Ahn, Man{\v{c}}al, Cheng,
  Blankenship, and Fleming}]{Engel-07}
\bibinfo{author}{G.~S. Engel}, \bibinfo{author}{T.~R. Calhoun},
  \bibinfo{author}{E.~L. Read}, \bibinfo{author}{T.-K. Ahn},
  \bibinfo{author}{T.~Man{\v{c}}al}, \bibinfo{author}{Y.-C. Cheng},
  \bibinfo{author}{R.~E. Blankenship}, \bibinfo{author}{G.~R. Fleming},
  \bibinfo{journal}{Nature} \bibinfo{volume}{446} (\bibinfo{year}{2007})
  \bibinfo{pages}{782--6}.
%Type = Article
\bibitem[{Shelly et~al.(2008)Shelly, Golovich, Dillman, and Beck}]{Shelly-08}
\bibinfo{author}{K.~R. Shelly}, \bibinfo{author}{E.~C. Golovich},
  \bibinfo{author}{K.~L. Dillman}, \bibinfo{author}{W.~F. Beck},
  \bibinfo{journal}{J. Phys. Chem. B} \bibinfo{volume}{112}
  (\bibinfo{year}{2008}) \bibinfo{pages}{1299--1307}.
%Type = Book
\bibitem[{Wolf(2007)}]{wolf2007}
\bibinfo{author}{E.~Wolf}, \bibinfo{title}{Introduction to the Theory of
  Coherence and Polarization of Light}, \bibinfo{publisher}{Cambridge
  University Press}, \bibinfo{year}{2007}.
%Type = Article
\bibitem[{Mazin et~al.(2012)Mazin, Bumble, Meeker, O'Brien, McHugh, and
  Langman}]{mazin2012}
\bibinfo{author}{B.~A. Mazin}, \bibinfo{author}{B.~Bumble},
  \bibinfo{author}{S.~R. Meeker}, \bibinfo{author}{K.~O'Brien},
  \bibinfo{author}{S.~McHugh}, \bibinfo{author}{E.~Langman},
  \bibinfo{journal}{Optics express} \bibinfo{volume}{20} (\bibinfo{year}{2012})
  \bibinfo{pages}{1503--1511}.
%Type = Article
\bibitem[{Mazin et~al.(2013)Mazin, Meeker, Strader, Szypryt, Marsden, van
  Eyken, Duggan, Walter, Ulbricht, Johnson et~al.}]{mazin2013}
\bibinfo{author}{B.~Mazin}, \bibinfo{author}{S.~Meeker},
  \bibinfo{author}{M.~Strader}, \bibinfo{author}{P.~Szypryt},
  \bibinfo{author}{D.~Marsden}, \bibinfo{author}{J.~van Eyken},
  \bibinfo{author}{G.~Duggan}, \bibinfo{author}{A.~Walter},
  \bibinfo{author}{G.~Ulbricht}, \bibinfo{author}{M.~Johnson}, et~al.,
  \bibinfo{journal}{Publications of the Astronomical Society of the Pacific}
  \bibinfo{volume}{125} (\bibinfo{year}{2013}) \bibinfo{pages}{1348--1361}.
%Type = Article
\bibitem[{Reiserer et~al.(2013)Reiserer, Ritter, and Rempe}]{reiserer2013}
\bibinfo{author}{A.~Reiserer}, \bibinfo{author}{S.~Ritter},
  \bibinfo{author}{G.~Rempe}, \bibinfo{journal}{Science} \bibinfo{volume}{342}
  (\bibinfo{year}{2013}) \bibinfo{pages}{1349--1351}.
%Type = Article
\bibitem[{Isoshima et~al.(1995)Isoshima, Isojima, Hakomori, Kikuchi, Nagai, and
  Nakagawa}]{isoshima1995}
\bibinfo{author}{T.~Isoshima}, \bibinfo{author}{Y.~Isojima},
  \bibinfo{author}{K.~Hakomori}, \bibinfo{author}{K.~Kikuchi},
  \bibinfo{author}{K.~Nagai}, \bibinfo{author}{H.~Nakagawa},
  \bibinfo{journal}{Review of scientific instruments} \bibinfo{volume}{66}
  (\bibinfo{year}{1995}) \bibinfo{pages}{2922--2926}.
%Type = Article
\bibitem[{Iranifam et~al.(2010)Iranifam, Segundo, Santos, Lima, and
  Sorouraddin}]{iranifam2010}
\bibinfo{author}{M.~Iranifam}, \bibinfo{author}{M.~A. Segundo},
  \bibinfo{author}{J.~L. Santos}, \bibinfo{author}{J.~L. Lima},
  \bibinfo{author}{M.~H. Sorouraddin}, \bibinfo{journal}{Luminescence}
  \bibinfo{volume}{25} (\bibinfo{year}{2010}) \bibinfo{pages}{409--418}.
%Type = Article
\bibitem[{Voeikov et~al.(2001{\natexlab{a}})Voeikov, Koldunov, and
  Kononov}]{voeikov4}
\bibinfo{author}{V.~L. Voeikov}, \bibinfo{author}{V.~V. Koldunov},
  \bibinfo{author}{D.~S. Kononov}, \bibinfo{journal}{Kinetics and Catalysis}
  \bibinfo{volume}{42} (\bibinfo{year}{2001}{\natexlab{a}})
  \bibinfo{pages}{606--608}.
%Type = Article
\bibitem[{Voeikov et~al.(2001{\natexlab{b}})Voeikov, Koldunov, and
  Kononov}]{voeikov5}
\bibinfo{author}{V.~L. Voeikov}, \bibinfo{author}{V.~V. Koldunov},
  \bibinfo{author}{D.~S. Kononov}, \bibinfo{journal}{Russian Journal of
  Physical Chemistry} \bibinfo{volume}{75} (\bibinfo{year}{2001}{\natexlab{b}})
  \bibinfo{pages}{1443–1448}.
%Type = Inproceedings
\bibitem[{van Wijk et~al.(2011)van Wijk, van~der Greef, and van
  Wijk}]{vanWijk2011}
\bibinfo{author}{E.~van Wijk}, \bibinfo{author}{J.~van~der Greef},
  \bibinfo{author}{R.~van Wijk}, in: \bibinfo{booktitle}{Journal of Physics:
  Conference Series}, volume \bibinfo{volume}{329}, \bibinfo{organization}{IOP
  Publishing}, p. \bibinfo{pages}{012021}.
%Type = Article
\bibitem[{Scholkmann et~al.(2011)Scholkmann, Cifra, Moraes, and
  de~Mello~Gallep}]{scholkmann2011}
\bibinfo{author}{F.~Scholkmann}, \bibinfo{author}{M.~Cifra},
  \bibinfo{author}{T.~Moraes}, \bibinfo{author}{C.~de~Mello~Gallep},
  \bibinfo{journal}{Journal of Physics: Conference Series}
  \bibinfo{volume}{329} (\bibinfo{year}{2011}) \bibinfo{pages}{012020}.
%Type = Article
\bibitem[{Goldberger et~al.(2002)Goldberger, Amaral, Hausdorff, Ivanov, Peng,
  and Stanley}]{goldberger2002}
\bibinfo{author}{A.~Goldberger}, \bibinfo{author}{L.~Amaral},
  \bibinfo{author}{J.~Hausdorff}, \bibinfo{author}{P.~Ivanov},
  \bibinfo{author}{C.~Peng}, \bibinfo{author}{H.~Stanley},
  \bibinfo{journal}{Proceedings of the National Academy of Sciences of the
  United States of America} \bibinfo{volume}{99} (\bibinfo{year}{2002})
  \bibinfo{pages}{2466--2472}.
%Type = Inproceedings
\bibitem[{Gu(1992)}]{Gu-92}
\bibinfo{author}{Q.~Gu}, in: \bibinfo{editor}{F.-A. Popp},
  \bibinfo{editor}{K.~H. Li}, \bibinfo{editor}{Q.~Gu} (Eds.),
  \bibinfo{booktitle}{Recent Advances in Biophoton Research and its
  Applications}, \bibinfo{publisher}{World Scientific},
  \bibinfo{address}{Singapore}, \bibinfo{year}{1992}, pp.
  \bibinfo{pages}{59--112}.
%Type = Inproceedings
\bibitem[{Gu(1995)}]{Gu-95}
\bibinfo{author}{Q.~Gu}, in: \bibinfo{editor}{L.~V. Beloussov},
  \bibinfo{editor}{F.-A. Popp} (Eds.), \bibinfo{booktitle}{Biophotonics --
  Non-equilibrium and Coherent Systems in Biology, Biophysics and
  Biotechnology}, \bibinfo{publisher}{Bioinform Services Co},
  \bibinfo{address}{Moscow}, \bibinfo{year}{1995}, pp.
  \bibinfo{pages}{116--135}.
%Type = Inproceedings
\bibitem[{Gu(1998)}]{Gu-98}
\bibinfo{author}{Q.~Gu}, in: \bibinfo{editor}{J.-J. Chang},
  \bibinfo{editor}{J.~Fisch}, \bibinfo{editor}{F.-A. Popp} (Eds.),
  \bibinfo{booktitle}{Biophotons}, \bibinfo{publisher}{Kluwer Academic
  Publishers}, \bibinfo{address}{Dordrecht}, \bibinfo{year}{1998}, pp.
  \bibinfo{pages}{299--321}.
%Type = Book
\bibitem[{Gu(2003)}]{gu2003}
\bibinfo{author}{Q.~Gu}, \bibinfo{title}{Radiation and bioinformation},
  \bibinfo{publisher}{Science press}, \bibinfo{year}{2003}.
%Type = Inproceedings
\bibitem[{Kun et~al.(2009)Kun, Liu, and Hun-Yu}]{Kun-09}
\bibinfo{author}{S.~Kun}, \bibinfo{author}{C.-L. Liu},
  \bibinfo{author}{J.~Hun-Yu}, in: \bibinfo{editor}{X.~Shen},
  \bibinfo{editor}{X.-L. Yang}, \bibinfo{editor}{X.-R. Zhang},
  \bibinfo{editor}{Z.-J. Cui}, \bibinfo{editor}{L.~J. Kricka},
  \bibinfo{editor}{P.~E. Stanley} (Eds.), \bibinfo{booktitle}{Proceedings of
  the 15th International Symposium on Bioluminescence and Chemiluminescence},
  \bibinfo{publisher}{World Scientific}, \bibinfo{address}{Singapore},
  \bibinfo{year}{2009}, pp. \bibinfo{pages}{63--6}.
%Type = Article
\bibitem[{Lozneanu and Sanduloviciu(2008)}]{Lozneanu-08}
\bibinfo{author}{E.~Lozneanu}, \bibinfo{author}{M.~Sanduloviciu},
  \bibinfo{journal}{Roman. Repts. Phys.} \bibinfo{volume}{60}
  (\bibinfo{year}{2008}) \bibinfo{pages}{885--98}.
%Type = Article
\bibitem[{van Wijk et~al.(2008)van Wijk, van Wijk, and Bajpai}]{Wijk-08}
\bibinfo{author}{E.~P.~A. van Wijk}, \bibinfo{author}{R.~van Wijk},
  \bibinfo{author}{R.~P. Bajpai}, \bibinfo{journal}{Indian J. Exp. Biol.}
  \bibinfo{volume}{46} (\bibinfo{year}{2008}) \bibinfo{pages}{345--52}.
%Type = Article
\bibitem[{Bajpai and Drexel(2008)}]{Bajpai-08}
\bibinfo{author}{R.~P. Bajpai}, \bibinfo{author}{M.~Drexel},
  \bibinfo{journal}{J. Acupunct. Meridian Stud.} \bibinfo{volume}{1}
  (\bibinfo{year}{2008}) \bibinfo{pages}{114--20}.

\end{thebibliography}

\end{document}